\input harvmac
\input amssym.def
\input epsf.tex
\def\N{{\cal N}}

\def\la{{\lambda}}

\def\H{{\cal H}}
\def\i{{\rm i}}
\def\g{{\frak g}}
\def\ie{{\it i.e.}}

\font\tenmsb=msbm10       \font\sevenmsb=msbm7
\font\fivemsb=msbm5       \newfam\msbfam
\textfont\msbfam=\tenmsb  \scriptfont\msbfam=\sevenmsb
\scriptscriptfont\msbfam=\fivemsb
\def\Bbb#1{{\fam\msbfam\relax#1}}

\def\Zop{{\Bbb Z}}

\def\bbbc{{\mathchoice {\setbox0=\hbox{$\displaystyle\rm C$}\hbox{\hbox
to0pt{\kern0.4\wd0\vrule height0.9\ht0\hss}\box0}}
{\setbox0=\hbox{$\textstyle\rm C$}\hbox{\hbox
to0pt{\kern0.4\wd0\vrule height0.9\ht0\hss}\box0}}
{\setbox0=\hbox{$\scriptstyle\rm C$}\hbox{\hbox
to0pt{\kern0.4\wd0\vrule height0.9\ht0\hss}\box0}}
{\setbox0=\hbox{$\scriptscriptstyle\rm C$}\hbox{\hbox
to0pt{\kern0.4\wd0\vrule height0.9\ht0\hss}\box0}}}}

\def\figin{\epsfcheck\figin}\def\figins{\epsfcheck\figins}
\def\epsfcheck{\ifx\epsfbox\UnDeFiNeD
\message{(NO epsf.tex, FIGURES WILL BE IGNORED)}
\gdef\figin##1{\vskip2in}\gdef\figins##1{\hskip.5in}
\else\message{(FIGURES WILL BE INCLUDED)}%
\gdef\figin##1{##1}\gdef\figins##1{##1}\fi}
\def\DefWarn#1{}
\def\figinsert{\goodbreak\midinsert}
\def\ifig#1#2#3{\DefWarn#1\xdef#1{fig.~\the\figno}
\writedef{#1\leftbracket fig.\noexpand~\the\figno}%
\figinsert\figin{\centerline{#3}}\medskip\centerline{\vbox{\baselineskip12pt
\advance\hsize by -1truein\noindent\footnotefont{\bf Fig.~\the\figno:} #2}}
\bigskip\endinsert\global\advance\figno by1}

\lref\gg{M.R. Gaberdiel, T. Gannon, {\it Boundary states for WZW
models}, Nucl. Phys. {\bf B639}, 471 (2002); {\tt hep-th/0202067}.}

\lref\bou{P. Bouwknegt, P. Dawson, A. Ridout, {\it D-branes on group
manifolds and fusion rings}, JHEP {\bf 0212}, 065 (2002);
{\tt hep-th/0210302}.}

\lref\fs{S. Fredenhagen, V. Schomerus, {\it Branes on group manifolds,
gluon condensates, and twisted K-theory}, JHEP 
{\bf 0104}, 007 (2001); {\tt hep-th/0012164}.}

\lref\mms{J. Maldacena, G. Moore, N. Seiberg, {\it D-brane instantons
and K-theory charges}, JHEP {\bf 0111}, 062 (2001);
{\tt hep-th/0108100}.}

\lref\braun{V. Braun, {\it Twisted K-theory of Lie groups}, 
{\tt hep-th/0305178}.}

\lref\dz{P. Di Francesco, J.-B. Zuber, {\it SU(N) lattice integrable 
models associated with graphs}, Nucl. Phys. {\bf B338}, 602 (1990).} 
 
\lref\mm{R. Minasian, G. Moore, {\it K-theory and Ramond-Ramond
charge}, JHEP {\bf 9711}, 002 (1997); 
{\tt hep-th/9710230}.}

\lref\wittenK{E. Witten, {\it D-branes and K-theory}, JHEP
{\bf 9812}, 019 (1998); {\tt hep-th/9810188}.}

\lref\kapustin{A. Kapustin, {\it D-branes in a topologically
nontrivial B-field}, Adv. Theor. Math. Phys. {\bf 4}, 127 (2000);
{\tt hep-th/9909089}.}

\lref\boum{P. Bouwknegt, V. Mathai, {\it D-branes, B-fields and twisted
K-theory}, JHEP {\bf 0003}, 007 (2000); {\tt hep-th/0002023}.}

\lref\fht{D. Freed, M. Hopkins, C. Teleman, {\it Twisted K-theory and
loop group representations}, {\tt math.at/0312155} and unpublished.}

\lref\ggone{M.R. Gaberdiel, T. Gannon, {\it The charges of a twisted
brane}, JHEP {\bf 0401}, 018 (2004); {\tt hep-th/0311242}.}

\lref\fgk{G. Felder, K. Gawedzki and A. Kupiainen, {\it The spectrum
of Wess-Zumino-Witten models}, 
Nucl. Phys. {\bf B 299}, 355 (1988).}

\lref\fgka{G. Felder, K. Gawedzki and A. Kupiainen, {\it Spectra of
Wess-Zumino-Witten models with arbitrary simple groups}, 
Commun. Math. Phys. {\bf 117}, 127 (1988).}

\lref\gew{D. Gepner, E. Witten, {\it String theory on group manifolds},
Nucl. Phys. {\bf B278}, 493 (1986).}

\lref\stanciu{S. Stanciu, {\it An illustrated guide to D-branes in
${\rm SU}_3$}, {\tt hep-th/0111221}.}

\lref\mss{K. Matsubara, V. Schomerus, M. Smedback,
{\it Open strings in simple current orbifolds}, 
Nucl. Phys. {\bf B626}, 53 (2002); {\tt hep-th/0108126}.}

\lref\gw{T. Gannon, M.A. Walton, {\it On fusion algebras and modular
matrices}, Commun. Math. Phys. {\bf 206}, 1 (1999); 
{\tt q-alg/9709039}.}  

\lref\intril{K. Intriligator, {\it Bonus Symmetry in Conformal
Field Theory}, Nucl. Phys. {\bf B332}, 541 (1990).}

\lref\sy{A.N. Schellekens, S. Yankielowicz, {\it Extended chiral
algebras and modular invariant partition functions}, Nucl. Phys. 
{\bf B327}, 673 (1989).}

\lref\fhssw{J. Fuchs, L.R. Huiszoon, A.N. Schellekens, C. Schweigert, 
J. Walcher, {\it Boundaries, crosscaps and simple currents},
Phys. Lett. {\bf B495}, 427 (2000); {\tt hep-th/0007174}.}

\lref\fsone{J. Fuchs, C. Schweigert, {\it Symmetry breaking boundaries
I. General theory}, Nucl. Phys. {\bf B558}, 419 (1999); 
{\tt hep-th/9902132}.}

\lref\fstwo{J. Fuchs, C. Schweigert, {\it Symmetry breaking boundaries
II. More structures; examples}, Nucl. Phys. {\bf B568}, 543 (2000);
{\tt hep-th/9908025}.}

\lref\as{A. Alekseev, V. Schomerus, {\it D-branes in the WZW model}, 
Phys. Rev. {\bf D60}, 061901 (1999); {\tt hep-th/9812193}.}

\lref\fss{J. Fuchs, B. Schellekens, C. Schweigert, {\it A matrix S for 
all simple current extensions}, Nucl. Phys. {\bf B473}, 323 (1996);}

\lref\bfs{L. Birke, J. Fuchs, C. Schweigert, {\it Symmetry breaking  
boundary conditions and WZW  orbifolds}, Adv. Theor. Math. Phys. 
{\bf 3}, 671 (1999); {\tt hep-th/9905038}.}

\lref\fss{J. Fuchs, B. Schellekens, C. Schweigert, {\it From Dynkin
diagram symmetries to fixed point structures},
Commun. Math. Phys. {\bf 180}, 39 (1996); {\tt hep-th/9506135}.} 

\lref\gaga{M.R. Gaberdiel, T. Gannon, {\it in preparation}.}

\lref\couch{N. Couchoud, {\it D-branes and orientifolds of SO(3)}, 
 JHEP {\bf 0203}, 026 (2002); {\tt hep-th/0201089}.}

\lref\pss{G. Pradisi, A. Sagnotti, Y.S. Stanev, 
{\it Planar duality in $SU(2)$ WZW models}, 
Phys. Lett. {\bf B354}, 279 (1995); {\tt hep-th/9503207}.}

\lref\pssa{G. Pradisi, A. Sagnotti, Y.S. Stanev, 
{\it The open descendants of non-diagonal SU(2) WZW models},
Phys. Lett. {\bf B356}, 230 (1995); {\tt hep-th/9506014}.}

\lref\pssb{G. Pradisi, A. Sagnotti, Y.S. Stanev, 
{\it Completeness conditions for boundary operators in 2D conformal
field theory}, Phys. Lett. {\bf B381}, 97 (1996);
{\tt hep-th/9603097}.}

\lref\fffs{G. Felder, J. Fr\"ohlich, J. Fuchs, C. Schweigert,
{\it The geometry of WZW branes}, J. Geom. Phys. {\bf 34}, 162 (2000);
{\tt hep-th/9909030}.}

\lref\bsn{V. Braun, S. Sch\"afer-Nameki, {\it Supersymmetric WZW models and
twisted K-theory of SO(3)}, {\tt hep-th/0403287}.}

\lref\fre{S. Fredenhagen, {\it D-brane charges on SO(3)}, {\tt hep-th/0404017}.}


\Title{\vbox{\baselineskip12pt
\hbox{hep-th/0403011}
\hbox{}}}
{\vbox{\centerline{D-brane charges on non-simply connected groups}}}
\smallskip
\centerline{Matthias R. Gaberdiel%
\footnote{$^\ast$}{{\tt gaberdiel@itp.phys.ethz.ch}}} 
\smallskip
\centerline{\it Institute for Theoretical Physics, ETH H\"onggerberg}
\centerline{\it CH-8093 Z\"urich, Switzerland}
\bigskip
\centerline{and}
\bigskip
\centerline{Terry Gannon%
\footnote{$^\star$}{{\tt tgannon@math.ualberta.ca}}}
\smallskip
\centerline{\it Department of Mathematical Sciences, University of
Alberta} 
\centerline{\it Edmonton, Alberta, Canada, T6G 2G1}\bigskip
\medskip
\vskip1.5cm
\centerline{\bf Abstract}
\bigskip
\noindent The maximally symmetric D-branes of string theory on the
non-simply connected Lie group SU$(n)/\Zop_d$ are analysed using
conformal field theory methods, and their charges are determined.
Unlike the well understood case for simply connected groups, the
charge equations do not determine the charges uniquely, and
the charge group associated to these D-branes is therefore in general
not cyclic. The precise structure of the charge group depends on some
number theoretic properties of $n$, $d$, and the level of the
underlying affine algebra $k$. The examples of 
SO$(3)=$SU$(2)/\Zop_2$ and SU$(3)/\Zop_3$ are worked out in detail,
and the charge groups for SU$(n)/\Zop_d$ at most levels $k$ are
determined explicitly.

\Date{February 2004}

\newsec{Introduction}

It is believed that the topological charges of D-branes are described
by some K-theory group \refs{\mm,\wittenK}. These charges constrain
the dynamics of D-branes as configurations that carry different
charges cannot decay into one another. For the case of the WZW models,
the relevant K-groups are believed to be certain twisted K-theory
groups \refs{\kapustin,\boum}.  

The D-brane charge groups are also calculable in terms of a
microscopic (conformal field theory) description of D-branes. Let us
denote by ${\cal B}$ the (finite) set of D-branes that preserve the
full affine symmetry $\g_k$ (possibly up to an automorphism). The
charge $q_a\in\Zop$ of the D-brane $a\in{\cal B}$ must then satisfy the
charge equation 
\refs{\fs}
\eqn\basic{
\dim(\lambda)\, q_a =
\sum_{b\in  {\cal B}} {\cal N}_{\la a}{}^b \, q_b\,.}
Here $\lambda\in P_+^k(\g)$ is an integrable highest weight
 of $\g_k$, $\dim(\lambda)$ is the Weyl dimension of the
corresponding finite dimensional representation of the horizontal
subalgebra $\bar\g$, and ${\cal N}_{\la a}{}^b$ are the NIM-rep
coefficients that describe the multiplicity (possibly zero) with
which the representation $\lambda$ appears in the open string spectrum 
of an open string that begins on the D-brane $a$, and ends on the
D-brane $b$. 

In the simplest situation (namely for the D-branes in the charge
conjugation modular invariant that preserve the full affine algebra
without any automorphism) the D-branes are labelled by the integrable
highest weight representations, $a=\mu$, and the NIM-rep 
${\cal N}_{\la \mu}{}^{\nu}$ agrees with the fusion rules
$N_{\la \mu}{}^{\nu}$. The charges $q_{\mu}$ are then (up to trivial
rescalings) uniquely determined by \basic, $q_{\mu} = \dim(\mu)$, and
they satisfy the charge equation modulo an integer $M$ 
\eqn\basicsun{
{\rm dim}(\la)\,{\rm dim}(\mu)=\sum_{\nu\in 
P_{+}^{k}}N_{\la\mu}{}^{\nu}\, {\rm dim}(\nu)\qquad({\rm mod}\ 
M)\,.}
The integer $M$ has been determined for all algebras and levels
\refs{\fs,\mms,\bou}, and it is given by the universal formula 
\eqn\Mdef{
M={k+h^\vee \over{\rm gcd}(k+h^\vee,L)}\,,}
where $h^\vee$ is the dual Coxeter number of the finite dimensional
Lie algebra $\bar{\g}$, and $L$ depends only on $\bar{\g}$. For the
case of su$(n)$ that shall mainly concern us in this paper,    
\eqn\Msun{
M=M_{{\rm SU}(n)}={n+k\over{\rm gcd}(n+k,L)}\,,}
where $L={\rm lcm}\{1,2,\ldots,n-1\}$. 

Since the coefficients ${\cal N}_{\la a}{}^b$ form a
(NIM-)representation of the fusion rules, any solution to \basic\ can
only be satisfied mod $M$, as is explained in \refs{\ggone}. In
general, however, the solutions of \basic\ will not be unique, and the
set of all solutions to \basic\ forms a $\Zop_M$-module $K$. [This is
to say, the sum of two solutions is a solution, and replacing $q_a$ by 
$lq_a$ (where $l\in\Zop_M$) also leads to a solution.] Since there are 
only finitely many D-branes (and since $M$ is a finite number), $K$
will only contain finitely many elements. 

The set of solutions $K$ is always an abelian group (under addition):
the identity element is the trivial charge solution ($q_a=0$ for all
$a$), and the inverse to the solution $q_a$ is the solution $-q_a$.
 This group is the  {\it charge group} of the
set of D-branes whose open string spectrum is described by  the
NIM-rep ${\cal N}_{\la a}{}^b$. Any finite abelian group is of the
form  
\eqn\Kgen{
K = \Zop_{M_1} \oplus \Zop_{M_2} \oplus \cdots \oplus \Zop_{M_l} \,.}
Apart from the trivial identification 
$\Zop_{a} \oplus \Zop_{b}=\Zop_{ab}$ whenever $a$ and $b$ are coprime,
this decomposition is unique. Since $K$ is a $\Zop_M$-module, all
$M_i$ must be factors of $M$. The charge group of the D-branes is
therefore always of this form.
\smallskip

Strings propagating on the simply connected group manifold $G$ are
described, in terms of conformal field theory, by the charge
conjugation modular invariant. For this theory, the charge group for
the D-branes that preserve the full affine symmetry has been
determined \refs{\fs,\mms,\bou}, and it is just $K=\Zop_M$, with $M$
given by \Mdef\ above. For the D-branes that only preserve the
affine symmetry up to an outer automorphism, the charge group has
recently also been found to be $K=\Zop_M$ \refs{\ggone}. These results
are in beautiful agreement with the K-theory calculations of
\refs{\fht,\braun}.

In this paper we want to determine the D-brane charge group for the
case of string theory on a group manifold that is not simply
connected. We shall only consider the case of quotient groups of the
simply connected group SU$(n)$, although many of our statements 
generalise directly to the other cases. Recall that the centre of
SU$(n)$ is $\Zop_n$. To each divisor $d>1$ of $n$, there is therefore
an associated compact connected (but not simply connected) Lie group 
$G=$ SU$(n)/\Zop_d$. Moreover, any compact connected Lie group with the
same Lie algebra su$(n)$ as SU$(n)$ will be of this form. The modular
invariant corresponding to $G$ is known \refs{\fgk,\fgka} and will be
reviewed in the following subsection. 

We shall only consider the `untwisted' D-branes in this paper, \ie\
the branes that preserve the full affine symmetry without any (outer)
automorphisms. These have been
implicitly constructed in \refs{\fhssw} (see also 
\refs{\fsone,\fstwo} for earlier work), but we will need to be more
explicit here, and shall review the relevant details below as well.  
As we shall see, unlike the situation in the simply connected case,
the charges here will not be uniquely determined by \basic, and $K$ is
therefore not a cyclic group in general. 

As will become apparent in the subsequent analysis, there are two
classes of solutions to \basic, which together generate all solutions. One 
solution (that we shall call
`untwisted') is characterised by the property that 
$q_0=1$, where $a=0$ describes the D0-brane of the underlying SU$(n)$
theory (that also defines a D-brane of SU$(n)/\Zop_d$). Provided that
$n(n+1)/d$ is even, this leads to a summand of $\Zop_{M_{{\rm SU}(n)}}$
in \Kgen. [This is for example illustrated by the example of
SU$(3)/\Zop_3$ that will be discussed very explicitly in 
section~2.] This charge solution simply measures the (rescaled)
D0-brane charge of the underlying SU$(n)$ theory. 

If $n(n+1)/d$ is odd on the other hand, the situation is different in
that the D0-brane of SU$(n)/\Zop_d$ (and therefore almost all D-branes
of SU$(n)/\Zop_d$) do not carry {\it any} non-trivial D0-brane charge
with respect to the underlying SU$(n)$ theory. In this case a
non-trivial solution with $q_0=1$ may still exist (although this need
not be the case), but it does not correspond to the D0-brane charge of
the underlying SU$(n)$ theory any more. In particular, it only gives
rise to a summand of $\Zop_{M^u}$ in \Kgen, where $M^u$ is typically either
$M^u=1$ or $M^u=2$, and at any rate does not grow with $k$ any more. This 
`pathological' behaviour already occurs for the simplest non-simply
connected Lie group SO$(3)=$ SU$(2)/\Zop_2$; this example will be
discussed very explicitly in section 2.3.  

In many cases the above untwisted solution is the unique solution to
\basic\ (in particular, this is the case provided that the level of
the affine Lie algebra $k$ is coprime to $d$), but there are also
situations where $q_0=1$ does not determine the full solution
uniquely. The ambiguity can then be described by `twisted' solutions
for which $q^{t}_0=0$. For example, if $d>2$ is prime (and $k$ is a
multiple of $d$), then there are precisely $d-1$ (linearly
independent) twisted solutions, each of which contributes a factor of    
\eqn\twissol{
\Zop_{M^t} \qquad \hbox{where} 
\qquad M^t = {\rm gcd}(d^\infty,n,M_{{\rm SU}(n)}) }
to \Kgen. If $d$ is composite (or $d=2$), the precise structure of the
twisted solutions is more complicated, but the general structure is
similar. The structure of the twisted solutions is nicely illustrated
by the example of SU$(3)/\Zop_3$.

We are not able to give a proof for these statements in all cases, but
we can give a nearly complete description when $d$ is a prime, and
when $d$ is composite but coprime to $M_{{\rm SU}(n)}$. We have also
studied a number of examples in detail, and the outlines of the general
picture are taking shape. One would expect that these  
charge groups should agree with the appropriate twisted K-theory
groups; it would be very interesting to confirm this by a direct
K-theory analysis.  
\medskip

The paper is organised as follows. In the remainder of this section we
introduce our notation and review the construction of the relevant
conformal field theories, as well as their untwisted D-branes. In
section~2 we illustrate our results by working out the examples of
SU$(3)/\Zop_3$ and SO$(3)=$ SU$(2)/\Zop_2$ explicitly. In section~3 we
collect some general observations and explain for which theories the
charge group is necessarily pathologically small. We also show
how to construct the unique (untwisted) solution provided that $k$ is
coprime to $d$. Finally we prove that the untwisted solution always
leads to $\Zop_{M_{{\rm SU}(n)}}$ provided that $d$ is coprime to 
$M_{{\rm SU}(n)}$ (and $n(n+1)/d$ is even so that the solution is not 
pathologically small). In section~4, we give a nearly complete
description of the charge group for the case when $d$ is
prime and comment on the generalisations to composite $d$.
Finally, section~5 contains our conclusions and conjectures about
general SU$(n)/\Zop_d$.

\subsec{A quick review of SU(n) and its simple current modular invariants}

String theory on SU$(n)$ can be described in terms of the
representation theory of the affine algebra
$\widehat{A}_{n-1}=\widehat{{\rm su}}(n)$ at the appropriate level $k$. Its
integrable highest weights $\lambda\in P_{+}^{k}({\rm su}(n))$ consists
of all $n$-tuples $(\lambda_{0};\lambda_{1},\ldots,\lambda_{n-1})$,
where each  $\lambda_{i}$ is a nonnegative integer, and 
$\sum_{i=0}^{n-1} \lambda_{i}=k$. When there can be no confusion
about the level $k$, we will often drop the redundant component
$\lambda_{0}$. These weights $\lambda\in P_{+}^{k}({\rm su}(n))$
parametrise the primary fields of the WZW model on SU$(n)$. 

This WZW model has a simple current $J$ of order
$n$, corresponding to the cyclic symmetry of the extended Dynkin
diagram, which sends the highest weight 
$\la=(\la_0;\la_1,\ldots,\la_{n-1})$ to
$J\la=(\la_{n-1}; \la_0,\la_1,\ldots,\la_{n-2})$. 
This permutation obeys  \refs{\sy,\intril}
\eqn\simcur{
S_{\lambda\, J^{j}\mu}=e^{2\pi \i j\,t(\lambda)/n}\,S_{\la\mu}}
for any $j$, where $t(\la)$ is its {\it n-ality}  
\eqn\nality{
t(\la)=\sum_{j=1}^{n-1} j\la_j\,.} 
Simple currents give rise to symmetries and gradings of fusion 
coefficients 
\eqn\scfus{\eqalign{
N_{J^i\lambda,J^j\mu}{}^{J^{i+j}\nu}=&\,N_{\la\mu}{}^\nu\cr
N_{\la\mu}{}^\nu\ne 0\ \Longrightarrow&\ t(\la)+t(\mu)=t(\nu)
\quad ({\rm mod}\ n)\,.}}

Suppose now that $d$ divides $n$, and write $d'=n/d$. The simple
current $J^{d'}$ that will play an important role in the following,
has then order $d$. We call $\varphi\in P_+^{k}$ a {\it fixed point of
order $m$} (with respect to $J^{d'}$) whenever $m$ divides $d$, and
$d/m$ is the smallest positive integer for which 
$J^{d'(d/m)}\varphi = J^{n/m}\varphi=\varphi$. Then 
the $J^{d'}$-orbit $\{J^{jd'}\varphi\}$ has cardinality  
$d/m$. Write $o(\varphi)$ for the order $m$ of
$\varphi$. Note that any solution $\varphi\in P_+^k({\rm su}(n))$ to
$J^{n/m}\varphi=\varphi$ looks 
like $\varphi=(\overline{\varphi},\ldots,\overline{\varphi})$ 
($m$ copies of $\overline{\varphi}$),  where 
$\overline{\varphi} =(\varphi_0;\ldots,\varphi_{n/m-1})\in 
P_+^{k/m}({\rm su}(n/m))$, and so
\eqn\nalfp{
t(\varphi)=\sum_{j=0}^{m-1}\sum_{i=0}^{n/m-1}({nj\over 
m}+i)\varphi_{i}={n\over m}{k\over m}{m(m-1)\over 
2}+m\,\overline{t}(\overline{\varphi})\,.}
For a given $n,k$ and $d$, $J^{d'}$ will have fixed points of order
$m$, when and only when $m$ divides gcd$(d,k)$.

For any divisor $d$ of $n$ with $d'=n/d$, define the matrix
$M[d']$ by   
\eqn\simcurM{
M[d']_{\la\mu}=
\sum_{j=1}^{d}\delta_{{d}}
\left(t(\la)+{d'\, j\, k' \over 2}\right)\,
\delta^{\mu\,J^{j d'}\la}\,,}  
where $k'=k+n$ if $k$ and $n$ are odd, and $k'=k$
otherwise. Furthermore, $\delta_y(x)=0$ if ${x\over y}\in{\Zop}$, and
$\delta_y(x)=1$ otherwise. Then $M[d']$ is a modular invariant
if and only if the product $(n-1)\, k\, d'$ is even \sy. For
instance, $M[n]=I$ is a modular invariant for any su($n$) level $k$;
for su(2), $M[1]$ is a modular invariant if and only if $k$ is
even. The modular invariant $M[d']$ corresponds to the WZW
model on the non-simply connected group SU$(n)/\Zop_{d}$
\refs{\fgk,\fgka}. For example, $M[1]$ for su$(2)$ at even level $k$ 
corresponds to the WZW model with group SO$(3)$.

\subsec{D-branes}

Next we need to describe the untwisted D-branes, and in particular,
their NIM-reps. In the following we shall consider SU$(n)/\Zop_d$,
where $d$ divides $n$. Suppose the level $k$ is fixed, and write 
$d'=n/d$ as before. In this subsection we assume for simplicity
that $M[d']$ defines a modular invariant, \ie\ that 
$(n+1)\, k\, d'$ is even.

We need to know the NIM-rep coefficients ${\cal N}_{\la a}{}^{b}$
corresponding to \simcurM. The easiest way to find these is through
the formula 
\eqn\psinim{
\N_{\la a}{}^{b}=\sum_{\mu\in{\cal 
E}}{\psi_{a\mu}S_{\la\mu}\psi_{b\mu}^{\ast}\over S_{0\mu}}\,,}
where $S$ is the Kac-Peterson modular $S$ matrix, and $\mu\in{\cal E}$ 
are the {\it exponents}. The general relation of $\psi$ with
the modular $S$ matrix for simple current extensions
is discussed in \refs{\bfs}; an expression for that $S$ is given in 
\refs{\fss}. The resulting general explicit expression for $\psi$ is 
too awkward to be directly helpful in \psinim; we will give special
instances of it throughout this paper. 

The fusion ring for SU$(n)$ at level $k$ is generated by the
fundamental weights $\Lambda_1,\ldots,\Lambda_{n-1}$. Since the
NIM-rep is a representation of the fusion ring, it suffices to
consider in \basic\ $\la$ equal to these fundamental weights
$\Lambda_i$. 

The exponents $\mu\in{\cal E}$ are the weights with 
$M[d']_{\mu\mu}\ne 0$. Write $f={\rm gcd}(d,k)$. Then
any order $o(\mu)$ must divide $f$. For the following it is useful to
distinguish two cases:\smallskip 

\noindent{{\bf Case A:}} Either $n$ or $k$ is odd, or either $n/d$ or
$k/f$ is even. Then $\mu\in P_+^k$ is an exponent if and only if $d$
divides $t(\mu)$. Such a $\mu$ will have multiplicity $o(\mu)$.
\smallskip

\noindent{{\bf Case B:}} Both $n$ and $k$ are even, and both $n/d$ and
$k/f$ are odd. Then the exponents come in two versions: either
$f/o(\mu)$ is even and $d$ divides $t(\mu)$ (in which case the
multiplicity of $\mu$ is $o(\mu)$); or $f/o(\mu)$ is odd and $d/2$
divides $t(\mu)$ (in which case the multiplicity of $\mu$ is
$o(\mu)/2$).\smallskip 

We will write `mult$(\mu)$' for this multiplicity (which will either be
$o(\mu)$ or $o(\mu)/2$), and let
$(\mu,i)$, $1\le i\le {\rm mult}(\mu)$, denote these exponents. 
When mult($\mu)=1$, then we shall usually abbreviate $(\mu,1)$ with $\mu$.
Write ${\cal E}={\cal E}_0\cup{\cal E}_{d/2}$ where ${\cal E}_i$ consists of
all exponents $\mu$ with $t(\mu)=i$ (mod $d$). Then ${\cal E}={\cal E}_0$
if and only if we are in Case A.

The boundary labels $a\in{\cal B}$ here correspond to pairs 
$([\nu],i)$, where $[\nu]=\{J^{j d'}\nu\}$ is the
$J^{d'}$-orbit of any weight $\nu\in P_{+}^{k}$, and where
$1\le i\le o(\nu)$. To simplify notation, we will write $[\nu,i]$ 
for $([\nu],i)$, and when $\nu$ has order $o(\nu)=1$, then we shall
usually write $[\nu]$ instead of $([\nu],1)=[\nu,1]$. 

Most entries of $\psi$ are easy to compute. When either $\mu$ or $\nu$
are not fixed points of $J^{d'}$, we find respectively 
\eqn\psimu{
\psi_{[\nu,i]\, \mu}={\sqrt{d}\over o(\nu)}\, S_{\nu\, \mu}\,,}
\eqn\psinu{
\psi_{[\nu]\, (\mu,i)}=\left\{
\eqalign{\sqrt{{d\over {\rm mult}(\mu)}}\, S_{\nu\, \mu}\quad
&{\rm if}\ \mu\in{\cal E}_0\cr 
0 \qquad\qquad\quad &{\rm if}\ \mu\in{\cal E}_{d/2}\,,}\right.}
for all $i$, where \psimu\ applies only if $o(\mu)=1$ (recall that
mult$(\mu)$ can equal $o(\mu)/2$ in Case B). More generally, we obtain
\eqn\psione{
\sum_{i=1}^{o(\nu)}\psi_{[\nu,i]\, (\mu,j)}=\left\{
\eqalign{
\sqrt{{d\over {\rm mult}(\mu)}}\, S_{\nu\, \mu}\quad
& {\rm if}\ \mu\in{\cal E}_0\cr 
0 \qquad\qquad\quad&{\rm if}
\ \mu\in{\cal E}_{d/2}\,,}\right.}
\eqn\psitwo{
\sum_{j=1}^{{\rm mult}(\mu)}\psi_{[\nu,i]\, 
(\mu,j)}=\left\{
\eqalign{{\sqrt{{\rm mult}(\mu) d}\over o(\nu)}\, S_{\nu\, \mu}\quad
&{\rm if} \ \mu\in{\cal E}_0\cr 
0\qquad\qquad\quad &{\rm if}\ \mu\in{\cal E}_{d/2}\,,}\right.}
for any boundary label $[\nu,i]$ and exponent $(\mu,j)$.

Now suppose $\nu$ is not a fixed point. We compute from \psinu\ 
and \psitwo\ that
\eqn\nimone{
\N_{\la\, [\nu]}{}^{[\nu',i]}=\sum_{\mu\in{\cal E}_0}
\sqrt{{d\over {\rm mult}(\mu)}}
S_{\nu\mu}{S_{\la\mu}\over S_{0\mu}}
{\sqrt{{\rm mult}(\mu) d}\over o(\nu')}S_{\nu'\mu}^\ast=\sum_{j=1}^{d/ 
o(\nu')}N_{\la\, \nu}{}^{J^{d'j}\nu'}\,,}
for any weight $\la\in P_+^k$ and boundary label $[\nu',i]$. More 
generally, for any weight $\la\in P_{+}^{k}$ and boundary labels 
$[\nu,i],[\nu',i']$, we obtain
\eqn\nimtwo{
\sum_{i=1}^{o(\nu)}\N_{\la\, [\nu,i]}{}^{[\nu',i']}=
\sum_{j=1}^{d/ 
o(\nu')}N_{\la\, \nu}{}^{J^{d'j}\nu'}\,.}
This does not give all NIM-rep coefficients, but it provides a very
useful constraint that can be exploited in general (see
section~3). For more specific examples (that will be discussed in
section~4) one can give simple explicit formulae for all NIM-rep
coefficients. In these cases one can then give a fairly complete
description of the D-brane charges.  More generally \gaga, each NIM-rep
coefficient $\N_{\la\,[\nu,i]}{}^{[\nu',i']}$ can always be expressed 
in terms  of various fusions for SU$(n/m)$ level $k/m$, for some $m$
dividing $f={\rm gcd}(d,k)$.

\newsec{The construction for ${\rm SU}(3)/\Zop_3$ and $SU(2)/\Zop_{2}$}

In order to illustrate our general results, which we shall give in
sections  3 and 4, let us first discuss the analysis 
for the case of ${\rm SU}(3)/\Zop_3$ and SU$(2)/\Zop_{2}$ in detail. 

\subsec{The SU$(3)$ situation without fixed points}

For SU$(3)/\Zop_{3}$, the situation is simplest if the level $k$ is
not divisible by $3$ since then the orbifold does not have any fixed 
points. (The situation when $3$ divides $k$ will be discussed in the
next subsection.) In this case,
\simcurM\ implies that the spectrum of the corresponding WZW model is
given by  
\eqn\suthreea{
\H = \bigoplus_{t(\lambda) = 0\, ({\rm mod}\ 3)} 
\H_\lambda \otimes \bar\H_{\lambda}^\ast \, \oplus \, 
\bigoplus_{t(\lambda) = 1\, ({\rm mod}\ 3)} 
\H_\lambda \otimes \bar\H_{J\lambda}^\ast \, 
\oplus \, 
\bigoplus_{t(\lambda) = 2\, ({\rm mod}\ 3)} 
\H_\lambda \otimes \bar\H_{J^2\lambda}^\ast\,,}
where $\lambda\in P_+^k({\rm su}(3))$ is an allowed highest weight
 for $\widehat{\rm su}(3)$ at level $k$, \ie\ 
$\lambda=(\lambda_1,\lambda_2)$ satisfies 
$\lambda_1+\lambda_2\leq k$. The quantity $t(\lambda)$ is
defined by  
\eqn\tdef{
t(\lambda) = \lambda_1 + 2 \lambda_2  \,,}
and $J$ is the simple current that acts on allowed highest weights as 
\eqn\Jdef{
J(\lambda_1,\lambda_2) =
(k-\lambda_1-\lambda_2,\lambda_1) \,.}
Since $k$ here is not divisible by $3$, none of the weights 
$\lambda\in P_+^k({\rm su}(3))$ are invariant under $J$.

The boundary states of this theory are labelled by the $J$-orbits in  
$P_+^k({\rm su}(3))$, and are denoted by $[a]$, where 
$a\in P_+^k({\rm su}(3))$. More explicitly, $[a]=[b]$ if and only if
$a=J^l b$, for some $l$. The corresponding NIM-rep is simply given by  
\eqn\suthreenima{
{\cal N}_{\lambda [a]}{}^{[b]} = \sum_{i=0}^2 
N_{\lambda a}{}^{J^i b} \,,}
where $N$ denotes the fusion rules of $\widehat{\rm su}(3)$ at level
$k$. We now make the ansatz that the charge of the D-brane
corresponding to $[a]$ is given by 
\eqn\suthreeb{
q_{[a]} = \dim_{\rm su(3)} (a)={(a_{1}+1)(a_{2}+1)(a_{1}+a_{2}+2)\over 2} \,.}
This definition is well-defined modulo $M_{{\rm  SU}(3)}$, 
\eqn\suthreec{
M_{{\rm  SU}(3)} = {k+3 \over {\rm gcd} (k+3,2) } \,,}
since 
\eqn\suthreed{
\dim(a) = \dim(Ja) \qquad ({\rm mod}\ M_{{\rm  SU}(3)}) \,,}
as was already observed in \refs{\fs}. By construction of 
$M_{{\rm  SU}(3)}$ (see \basicsun), we have 
\eqn\suthreee{
\dim(\lambda)\, \dim(a) = \sum_{b\in P_+^k({\rm su}(3))} \,
N_{\lambda a}{}^{b} \, \dim(b) \qquad  ({\rm mod}\ M_{{\rm  SU}(3)})
\,.} 
It then follows that the right hand side of \basic\ equals 
\eqn\suthreef{\eqalign{
\sum_{[b]} {\cal N}_{\lambda [a]}{}^{[b]} \, q_{[b]} & = 
\sum_{[b]} \sum_{i=0}^2 
N_{\lambda a}{}^{J^i b} \, \dim(b) \cr
& = \sum_{[b]} \sum_{i=0}^2 
N_{\lambda a}{}^{J^i b} \, \dim(J^i b) \qquad
({\rm mod}\ M_{{\rm  SU}(3)}) \cr
& = \sum_{b\in P_+^k({\rm su}(3))} \,
N_{\lambda a}{}^{b} \, \dim(b)  \cr
& = \dim(\lambda) \, q_{[a]} \qquad 
({\rm mod}\ M_{{\rm  SU}(3)}) \,.}}
Thus the ansatz \suthreeb\ solves \basic\ with $M=M_{{\rm  SU}(3)}$.
It is unique (as we explain in the next section), so the twisted
charges $q^{t}_{[a]}$ are all $0$, with $K=\Zop_{M}$.

\subsec{The SU(3) situation with fixed points}

The analysis is more complicated when $k$ is divisible by $3$, since  
$P_+^k({\rm su}(3))$ contains then a highest weight 
$\phi =(k/3,k/3)$ for which $J \phi = \phi$. In this case the
spectrum of the theory is given by 
\eqn\suthreeba{
\H = \bigoplus_{[\lambda], t(\lambda) = 0\, ({\rm mod}\ 3)} 
\Bigl( \H_\lambda \oplus \H_{J\lambda} \oplus 
\H_{J^2\lambda} \Bigr) \otimes 
\Bigl( \bar\H_\lambda^\ast \oplus \bar\H_{J\lambda}^\ast \oplus 
\bar\H_{J^2\lambda}^\ast \Bigr) \,
\oplus
3 \, \H_\phi \otimes \bar\H_\phi \,,}
where the first sum is over all $J$-orbits $[\lambda]$ with
$\lambda\ne\phi$ for which $t(\lambda)=0$ $({\rm mod}\ 3)$. As before,
some boundary states are labelled by the $J$-orbits $[a]$ with 
$a\ne \phi$, but now there are in addition $3$ boundary states
$[\phi,i]$ with $i=1,2,3$  associated to the fixed point
$\phi$. In addition to the NIM-rep coefficients \suthreenima\ we now
have \refs{\gg}
\eqn\suthreebb{
{\cal N}_{\lambda [a]}{}^{[\phi,i]} =
N_{\lambda a}{}^{\phi} }
as well as 
\eqn\suthreebc{ 
{\cal N}_{\lambda [\phi,i]}{}^{[\phi,j]} =  \left\{ 
\eqalign{ 
{1\over 3} N_{\lambda \phi}{}^{\phi} \qquad \qquad \quad & 
\hbox{if $N_{\lambda \phi}{}^{\phi} =0$ $({\rm mod}\ 3)$} \cr 
{1\over 3} (N_{\lambda \phi}{}^{\phi} -1) + \delta^{ij} 
\qquad & 
\hbox{if $N_{\lambda \phi}{}^{\phi} =1$ $({\rm mod}\ 3)$} \cr 
{1\over 3} (N_{\lambda \phi}{}^{\phi} +1) - \delta^{ij} 
\qquad & 
\hbox{if $N_{\lambda \phi}{}^{\phi} =2$ $({\rm mod}\ 3)$.}}
\right.}
For the following it is useful to note that 
$N_{\lambda \phi}{}^{\phi}=0$ if either $\lambda_1=2$ (mod 3) or 
$\lambda_1\ne \lambda_2$ $({\rm mod}\ 3)$.  

We now want to construct three independent solutions to \basic. The
first solution is associated to the `untwisted' D0-brane charge. To
this end we define as before
\eqn\suthreebd{ 
q_{[a]} = \dim(a) \,.}
In order to analyse whether this ansatz (together with a suitable
formula for $q_{[\phi,i]}$ that we shall deduce in the following) 
solves \basic, we observe that it is sufficient to check 
\basic\ for $\lambda$ equalling one of the two fundamental
weights $(1,0),(0,1)$ of ${\rm su}(3)$. Let us first consider \basic\ for
the situation where $a$ denotes a boundary $[a]$ with $a\ne\phi$. Then
the analysis of the previous subsection shows that the above charges
solve \basic\ modulo  $M_{{\rm  SU}(3)}$ provided that 
\eqn\suthreebf{ 
\sum_{i=1}^{3} q_{[\phi,i]} = \dim(\phi) \qquad
({\rm mod}\ M_{{\rm  SU}(3)}) \,.}
On the other hand if $a=[\phi,i]$, \basic\ is solved for 
$\lambda=(1,0)$ modulo $M_{{\rm  SU}(3)}$ provided that 
\eqn\suthreebg{
3 \, q_{[\phi,i]} = \dim((k/3+1,k/3)) \qquad 
({\rm mod}\ M_{{\rm  SU}(3)}) \,,}
with an equivalent relation for $\lambda=(0,1)$. We can solve  
\suthreebg\ by defining 
\eqn\suthreebe{ 
q_{[\phi,i]} = {1\over 3} \dim((k/3+1,k/3)) 
+ l_i {M_{{\rm  SU}(3)} \over 3} \,, }
where $l_i\in\Zop$ is defined modulo $3$ (given that $q_{[\phi,i]}$ 
is only defined modulo $M_{{\rm  SU}(3)} $). Since
$\dim((a,b))=(a+1)(b+1)(a+b+2)/2$, and $3$ divides $k$, the right hand
side of \suthreebe\ is indeed an integer. It therefore only remains to 
check whether this ansatz solves also \suthreebf. Using the above
dimension formula one finds that this is the case provided that 
\eqn\suthreeh{
\sum_{i=1}^{3} l_i = (-1)^k \qquad
({\rm mod}\ 3) \,.}
It is always possible to choose the $l_i$ in this manner, and thus we
have shown that \suthreebd\ and \suthreebe\ define a solution to
\basic\ with $M=M_{{\rm  SU}(3)}$. 

It is clear from the above discussion that this solution is not unique
(since only the sum of the $l_i$ is determined). If we subtract two
consistent solutions from one another, we obviously obtain another
consistent solution. The resulting solution can be interpreted as
measuring the `twisted' charges of the orbifold since $q^{t}_{[a]}=0$
for all non-fixed points $a$. If we make this ansatz, the analogue of
\suthreebg\ now implies that  
\eqn\suthreej{
3\, q^{t}_{[\phi,i]} = 0 \qquad ({\rm mod}\ M^{t}) \,,}
and thus that $M^{t}=3$. Furthermore, the analogue of \suthreebf\
gives that 
\eqn\suthreek{
\sum_{i=1}^{3}  \, q^{t}_{[\phi,i]} = 0 \qquad ({\rm mod}\ 3) \,.}
It is clear that there are two independent solutions to these
equations: we can choose $q^{t}_{[\phi,1]}$ and
$q^{t}_{[\phi,2]}$ independently, and $q^{t}_{[\phi,3]}$ is then
determined by \suthreek. Thus the charge vectors are $3$-dimensional
in this case, and the total charge group is 
\eqn\totalcharge{
K = \Zop_{M_{{\rm  SU}(3)}} \oplus \Zop_3 \oplus \Zop_3 \,.}

\subsec{The example of ${\rm SO}(3)$}

The final phenomena which can occur is well-illustrated by
${\rm SO}(3)$, the simplest example of a non-simply connected Lie
group. Let us recall from the work of
\refs{\gew,\fgk,\fgka} that the  ${\rm SO}(3)$ WZW model only exists
for even $k$, and that its spectrum is given by
\eqn\sothree{
\H = \bigoplus_{j=0}^{k/2} \H_{2j} \otimes \bar\H_{2j} \oplus
\bigoplus_{j=1}^{k/2 } \H_{2j-1} \otimes \bar\H_{k+1-2j} \,,}
provided that $4$ does not divide $k$ (what we call `Case B' in 
section~1.2), and by 
\eqn\sothreea{
\H = \bigoplus_{j=0}^{k/4-1} 
\Bigl(\H_{2j} \oplus \H_{k-2j} \Bigr) \otimes 
\Bigl(\bar\H_{2j} \oplus \bar\H_{k-2j} \Bigr) \,
\oplus
2\, \H_{k/2}\otimes \bar\H_{k/2} \,,}
provided that $4$ does divide $k$ (`Case A'). Here $\H_j$ is
the highest weight representation of the affine  
$\widehat{{\rm su}}(2)$ algebra at level $k$ with highest weight $j$;
we choose the convention that $j$ is integral so that the vector
representations of ${\rm su}(2)$ are described by even $j$, while the
spinor representations correspond to odd $j$. 

It is easy to read off from \sothree\ and \sothreea\ that the theory
has $n=k/2+2$ 
exponents, \ie\ that there are $n$ different (untwisted) Ishibashi
states, and hence also $n$ different (untwisted) boundary states. For 
the usual reason, it is enough to consider the $2$-dimensional fundamental
representation of ${\rm su}(2)$. Its
NIM-rep matrix $\N_{1}$ is given by the $D_n$ Dynkin diagram
\refs{\dz} (the D-branes of this theory were first constructed in a
series of papers in \refs{\pss,\pssa,\pssb}):
\ifig\sunpic{The NIM-rep graph for su$(2)$ corresponding to the 
${\rm SO}(3)$ modular invariant $D_n$ for $n=k/2+2=9$.}
{\epsfxsize3.5in\hskip.2cm\epsfbox{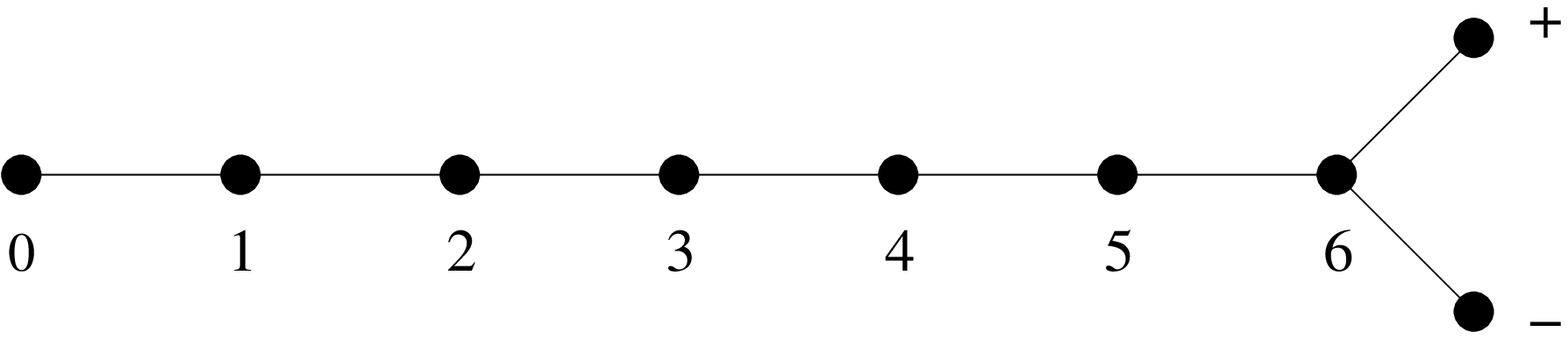}}
Let us label the boundary states starting from the left by
$a=0,\ldots,n-3$, with the two boundary states at the tail being
denoted by $a=\pm$. Then the charges, which we shall denote as $q_l$,
$l=0,\ldots,n-3$ as well as $q_\pm$,  have to satisfy the relations 
\eqn\chargesothreea{\eqalign{
2 \, q_0 & = q_1 \cr 
2 \, q_l & = q_{l-1} + q_{l+1} \qquad\qquad  l=1,\ldots, n-4 \cr
2 \, q_{n-3} & = q_{n-4} + q_+ + q_- \cr
2 \, q_+ & = q_{n-3} \cr
2 \, q_- & = q_{n-3} \,.}}
The first two equations imply that 
\eqn\chargessuthreeb{
q_l = (l+1)\, q_0 \,, \qquad l=0,\ldots, n-3\,.}
The third last equation then gives that 
\eqn\chargessuthreec{
q_+ + q_- = (n-1)\, q_0 \,,}
while the sum of the last two equations is $2(q_+ +q_-) = 2(n-2)q_0$.
By subtracting this equation from twice \chargessuthreec, it follows
that 
\eqn\chargessuthreed{
2\, q_0 =0 \qquad ({\rm mod}\ M) \,.}
For the untwisted charge component, $q_0=1$, and $M=2$. If $k$ is
divisible by $4$, $n$ is even, and $q_++q_-=1$, as well as  
$2 q_+= 2 q_-=0$ mod $2$. Thus the untwisted charge solution is 
\eqn\chargessuthreee{
\hbox{$k$ divisible by $4$:} \qquad \qquad 
q_l = \left\{ \eqalign{ 1 \qquad & \hbox{$l$ even} \cr
                        0 \qquad & \hbox{$l$ odd} } \right.
\qquad\qquad
q_+ = 1\,, \qquad q_- = 0\,.}
This charge assignment is not unique since we could have also chosen
$q_+=0$ and $q_-=1$. Thus we have a second (twisted) charge solution
with $M^t=2$ and 
\eqn\chargessuthreeeb{
\hbox{$k$ divisible by $4$:} \qquad \qquad 
q^t_l=0 \,, \qquad q^t_+=1 \,,\qquad q^t_-=1 \,.}
The corresponding charge group is therefore
\eqn\totalchargeone{
\hbox{$k$ divisible by $4$:}\qquad\qquad
K = \Zop_2 \oplus \Zop_2 \,.}
On the other hand, if $k$ is not divisible by $4$, $n$ is odd, and 
$2 q_+= 2 q_-=q_0$ mod $2$ with $q_+ + q_- = 0$ mod $2$. In this case 
the untwisted solution is (with $M=2$)
\eqn\chargessuthreef{
\hbox{$k$ not divisible by $4$:} \qquad
q_l = \left\{ \eqalign{ 1 \qquad & \hbox{$l$ even} \cr
                        0 \qquad & \hbox{$l$ odd} } \right.
\qquad
q_+ = {1\over 2} \qquad q_- = - {1\over 2} \,.}
Note that $3q_l=q_l$, and that $3q_+=-{1\over 2}$ as well as 
$3q_-={1\over 2}$. Thus this solution encompasses both solutions to
$q_++q_-=0$, and there is therefore no separate twisted solution --- 
\ie\ the twisted solution (also given by \chargessuthreeeb\ and
corresponding to  
$M^t=2$) is redundant.
Since we now have half-integer charges in \chargessuthreef, the
resulting charge group is in this case
\eqn\totalchargetwo{
\hbox{$k$ not divisible by $4$:}\qquad\qquad
K = \Zop_4 \,.}
The fractional charges \chargessuthreef\ require that the charge group 
$K$ in \totalchargetwo\ be formed from the group extension of 
$\Zop_{M}=\Zop_{2}$ by $\Zop_{M^t}=\Zop_{2}$, rather than their direct
sum as in \totalchargeone. Note that if $k$ is divisible by $4$,
the modular invariant is of Case A, while if $k$ is not divisible by
$4$, we are in Case B. 

This is different to what one may have expected based on the analysis
of \refs{\stanciu}, or what seems to be suggested in \refs{\mss}. In
particular, the above charge groups do not grow with $k$. This result
has also a simple quasi-geometric interpretation. As is explained in
\refs{\fffs,\mss,\couch} (see also \refs{\as}), the brane
corresponding to spin $l$ for SO$(3)$ is the $\Zop_2$ invariant
superposition of the SU$(2)$ brane corresponding to $l$, and the
SU$(2)$ brane corresponding to $k-l$. The D0-brane charge of these two
branes are $q_l=l+1$ and $q_{k-l}=k-l+1$, respectively.  The
configuration that appears in SO$(3)$ therefore has total D0-brane
charge $q=l+1+k-l+1=k+2$, and this vanishes since the charge group of
SU$(2)$ is $\Zop_{k+2}$. The 
only D-branes that carry untwisted D0-brane charge are therefore the
branes labelled by $a=\pm$, that correspond to the $\Zop_2$-invariant
2-cycle. These branes carry the same untwisted D-brane charge (namely 
$(k+2)/2$). Thus the D-branes of SO$(3)$ only see a $\Zop_2$ subgroup
of the untwisted D-brane charge group. (Confusingly, this is what is
measured by the `twisted' charge solutions above!)

More formally, the superpositions of boundary states that are relevant
for SO$(3)$ are related by the action of the simple current $J$. In
SU$(2)$, the charges of the brane associated to a representation $j$
is simply $\dim(j)=j+1$, and the SO$(3)$ invariant configurations do
not carry any charge since 
\eqn\sothreep{
\dim(Jj) = - \dim(j) \qquad ({\rm mod}\ M_{{\rm SU}(2)} ) \,.}
In fact, since $J$ is a simple current, this property already follows 
from the fact that $\dim(J0)=-1$ (mod $M_{{\rm SU}(2)}$).

Incidentally, it is intriguing that the charge groups for SO(3) level
$k$ match exactly the centre of the associated $D_n$ diagram
(here $n=k/2+2$).\footnote{$^\star$}{We thank Mark Walton for pointing
this out to us.} Likewise, the diagonal SU(2) level $k$ theory has
charge group $\Zop_{k+2}$, which matches the centre of the associated
$A_n$ diagram (here $n=k+1$). Easy calculations confirm that the SU(2)
exceptional models associated to $E_6$, $E_7$ and $E_8$ have charge 
groups $\Zop_3$, $\Zop_2$ and $\Zop_1$, respectively, which again
equal their centres. We do not have an explanation for this exact
matching between charge groups and centres for the SU(2) theories,
which seems to deepen the mysterious A-D-E correspondence here.

\newsec{General results}

Let us now study the D-brane charges and the charge groups in the
general case whose NIM-rep coefficients were at least partially
described in section~1.2. 

Recall from the introduction that any solution to \basic\ must have 
order dividing $M_{{\rm SU}(n)}$, so in particular both the order of
the untwisted and twisted solution, $M^u$ and $M^t$, must divide
$M_{{\rm SU}(n)}$. This follows by the same argument given in \ggone,
since the NIM-rep here is again a representation of the fusion ring.

First we want to analyse in which situations
the untwisted charge group is pathologically small, as was the case
for SO$(3)$ above.

\subsec{Characterising the pathological cases}

As for any simple current, $\N_{J^{d'}0}$ will be a permutation
matrix. From \nimtwo\ with $\la=J^{d'}0$, we see that 
$\N_{J^{d'}0,[\nu,i]}{}^{[\nu',i']}=0$ unless the $J^{d'}$-orbits
$\langle J^{d'}\rangle\nu$ and $\langle J^{d'}\rangle \nu'$ are
equal. Thus  
\eqn\Jdprime{
\N_{J^{d'}0,[\nu,i]}{}^{[\nu',i']}=
\delta_{\nu,\nu'}\delta_{i',\pi_\nu i}\,,} 
where $\pi_\nu$ is a permutation of the indices $\{1,\ldots,o(\nu)\}$. 
Putting  $\la=J^{d'}0$ and $a=[0]$ into \basic, we obtain the
important condition that for the untwisted solution with $q_{[0]}=1$ 
\eqn\cond{
{\rm dim}(J^{d'}0)=1\qquad ({\rm mod}\ M)\,.}
[Note that the integer $M$ that appears in \cond\ need not be the
order $M^u$ of the corresponding solution: we have normalised the
untwisted solution so that $q_{[0]}=1$, but fractional charges are
possible for fixed points, as was the case for the example of SO$(3)$
that was considered in section~2.3.]

\noindent Now, the simple current $J$ of su(n) level $k$ obeys 
\eqn\dimJ{
{\rm dim}(J0)= \prod_{j=1}^{n-1}{k+n-j\over j}=
(-1)^{n+1}\qquad ({\rm mod}\ M_{{\rm SU}(n)})}
and thus 
\eqn\dimJtwo{
{\rm dim}(J^{d'}0)=(-1)^{(n+1)d'}\qquad ({\rm mod}\ 
M_{{\rm SU}(n)})\,.}
It therefore follows that $M$ can be at most $2$ if $(n+1)d'$
is odd. The charge group for these theories will therefore be 
pathologically small (just as for SO$(3)$), and we shall sometimes
refer to them as being {\it pathological}. Incidentally, the theories 
for which $(n+1)d'$ is odd are precisely those for which the partition
function $M[d']$ is only modular invariant for $k$ even. From a
geometrical point of view this constraint on $k$ reflects some
obstruction in the definition of the Wess-Zumino term \refs{\fgka}.  

We observe that Case B always corresponds to $(n+1)d'$ odd, and will
therefore always exhibit the above pathological behaviour. On the
other hand, Case A may or may not be pathological, and it can only 
be pathological if $d$ is even.

We can say more, by studying $\N_{J^{d'}0}$ further. Consider first 
Case A in section~1.2, \ie\ either $n$ or $k$ is odd, or 
either $n/d$ or $k/f$ is even. Then one can show (by
considering their eigenvalues $S_{J^{d'}0\,\mu}/S_{0\,\mu}$) 
that in fact the matrix $\N_{J^{d'}0}$ is the identity matrix. Thus in
Case A, any charge solution (untwisted or twisted) can only be
satisfied mod $M$, where 
\eqn\condtw{
{\rm dim}(J^{d'}0)=1\qquad ({\rm mod}\ M)\,,}
and $M=M^u$ or $M=M^t$. So in particular, if Case A is pathological,
any solution to \basic\ has order at most $2$, and thus the total
charge group is a direct sum of $\Zop_2$'s. (An example of this
situation is SO$(3)$ for $k$ divisible by $4$.)

In Case B, the eigenvalues $S_{J^{d'}0\,\mu}/S_{0\,\mu}$ are all 
$\pm 1$, and therefore $\N_{J^{d'}0}$ necessarily has order 2. By
counting $-1$'s, we find that in \Jdprime\ $\pi_\nu(i)=i$ for all
boundary labels $[\nu,i]$, except for a number exactly equal to the
number of boundary labels $[\nu,i]$ for which  $f/o(\nu)$ is odd and
divides $\overline{t}(\overline{\nu})$. Thus it can be expected (but
we have not proven this) that the permutation $\pi_\nu$  
is nontrivial for precisely those $\nu$ and no others, 
in which case $\pi_\nu$ is order 2 without fixed points. In any case,
\condtw\ will not hold in general in Case B. (For example, it does not
hold for the case of SO$(3)$ when $4$ does not divide $k$.)

For completeness we also give the corresponding simple current
dimensions for all other groups:

For $B_n$: dim$(J0)=-1$ mod $M_{B}$ \refs{\ggone}; 

For $C_n$: dim$(J0)=(-1)^{n(n+1)/2}$ mod $M_{C}$;

For $D_n$: dim$(J_v0)=1$, and 
dim$(J_{s}0)={\rm dim}(J_{c}0)=(-1)^{n(n-1)/2}$ all taken 
mod $M_D$;

For $E_6$: dim$(J0)=1$ mod $M_{E_6}$; and finally

For $E_7$: dim$(J0)=-1$ mod $M_{E_7}$.

\noindent It follows from these results, that for example the D-brane
charges of B$_n/\Zop_2$ and E$_7/\Zop_2$ will be
pathologically small, and similarly for $C_n$ (provided that
$n(n+1)/2$ is odd) and $D_n$ (provided that $n(n-1)/2$ is odd and the
quotient group involves $J_c$ or $J_s$).

\subsec{The complete solution for $k$ coprime to $d$}

The untwisted solution is characterised by setting $q_{[0]}=1$. 
Then \nimone\ and \basic\ with $a=[0]$ require 
\eqn\chargenfp{
q_{[\la]}={\rm dim}(\la)}
for any non-fixed point $\la\in P_{+}^{k}$ of $J^{d'}$ (\ie\
$o(\la)=1$). If there is more than one solution for which
\chargenfp\ is satisfied, we can consider their difference, which
defines then a `twisted' solution $q^{t}_a$ with 
\eqn\twchnfp{
q^{t}_{[\la]}=0\,,}
for any $\la$ that is not a fixed point of $J^{d'}$. Note that these
results apply regardless of the value of $M$ or $M^{t}$. More
generally, we find from \nimtwo\ the conditions  
\eqn\chargesum{
\sum_{i=1}^{o(\la)}q_{[\la,i]}={\rm dim}(\la)}
and
\eqn\twchsum{
\sum_{i=1}^{o(\la)}q^{t}_{[\la,i]}=0\,,}
where $\la\in P_{+}^{k}$ is any weight.

We will use these equations in the following sections. For
now we observe that {\it they provide an immediate and complete
solution to the situation where there are no fixed points.}  This
occurs precisely when gcd$(k,d)=1$. This solution is given by
\chargenfp\ with $M=M_{{\rm SU}(n)}$ (if $(n+1)d'$ is even) or 
$M={\rm gcd}(2,M_{{\rm SU}(n)})$ (if $(n+1)d'$ is odd).
It is the unique solution, in the sense that any other solution 
is a multiple of this one. For this $M$, \chargenfp\ is well-defined, 
and \basic\ follows from \basicsun. On the other hand, the twisted 
charges must all vanish. Thus the charge group here is $K=\Zop_{M}$.
This analysis also applies directly to the other affine algebras.

The condition gcd$(k,d)=1$ is satisfied for example for 
SU(2) when the level $k$ is odd. The NIM-rep is then given by the
`tadpole graph' of \refs{\dz}. As is well known, this does not
define a modular invariant (since $k$ has to be even for the SO$(3)$
theory) and $M=1$ here (\ie\ all charges, twisted and untwisted, are 
trivial). The same will happen whenever $M[d']$ is not a modular 
invariant. In a sense this is the reason why the D-brane charges for
SO$(3)$ behave rather atypically: the simple general solution never
arises for the SO$(3)$ theory. (The actual analysis for SO$(3)$ was
described in detail in the last section.) On the other hand, there are 
obviously many modular invariants for which the condition gcd$(k,d)=1$ 
is satisfied.

\subsec{Some generalisations}

We can generalise this construction as follows. For simplicity assume
that $(n+1)d'$ is even so that $\dim(J^{d'}0)=1$ mod 
$M_{{\rm SU}(n)}$. (Otherwise, as is explained in section 3.1, the
charge group is anyway pathologically small.) We claim that a (in
general fractional) solution to \basic, taken modulo
$\widehat{M}=M/{\rm gcd}(d,M)$, where $M=M_{{\rm SU}(n)}$, is given by    
\eqn\chafp{
q_{[\mu,i]}={{\rm dim}(\mu)\over o(\mu)}\,,} 
for all $1\le i\le o(\mu)$ and all weights $\mu\in P_+^k$. 
(Note that this solution reduces to \chargenfp\ when $\mu$ is not a
fixed point.) Verifying that \chafp\ does indeed satisfy \basic\
amounts to showing  that
\eqn\verif{
{\rm dim}(\la)\,{\rm 
dim}(\mu)=o(\mu)\sum_{[\nu]}\sum_{i=1}^{o(\nu)}
\N_{\la\,[\mu,j]}{}^{[\nu,i]} \,
{{\rm dim}(\nu)\over o(\nu)}\qquad \left({\rm mod}\ 
{\rm gcd}(M,o(\mu)\widehat{M})\right)\,.}
In fact, we can show that this congruence holds mod $M$: by \nimtwo\ 
(or rather the transpose of this equation), the right-hand side
becomes 
\eqn\veriftwo{o(\mu)\sum_{[\nu]}\sum_{j=1}^{d/o(\mu)}
N_{\la\,J^{d'j}\mu}{}^{\nu}
{{\rm dim}(\nu)\over 
o(\nu)}=\sum_{[\nu]}\sum_{j=1}^{d/o(\nu)}
N_{\la\,\mu}{}^{J^{d'j}\nu}
{{\rm dim}(\nu)}=\sum_{\nu}
N_{\la\,\mu}{}^{\nu}{\rm dim}(\nu)\;\;({\rm mod}\ {M})\,,}
using  \scfus\ twice, as well as the fact that 
dim$(J^{d'j}\nu)={\rm dim}(\nu)$ mod $M$. Of course, \basicsun\ then
implies that the right-side of \veriftwo\ equals the left-side of
\verif\ mod $M$. 

Now, a very common situation is when $d$ is coprime to 
$M_{{\rm SU}(n)}$.  
This happens precisely when, for each prime $p$ dividing $d$, the
largest power  $p^{\beta}$ dividing $k+n$, obeys $p^{\beta}<n$. In
this case, $d$ is invertible mod $M_{{\rm SU}(n)}$ and since $o(\mu)$
divides $d$, the charges \chafp\ are manifestly integers. The
resulting charge for $[0]$ equals then $q_{[0]}=1$, and \chafp\
defines an untwisted solution mod $M_{{\rm SU}(n)}$, the maximal
possible value. 

However, when $d$ and $M_{{\rm SU}(n)}$ are not coprime, this solution
will not describe the most general untwisted solution (up to the
ambiguity described by a  twisted solution), because rescaling it by
$d$ leads to $q_{[0]}={\rm gcd}(d,M_{{\rm SU}(n)})$, not
$q_{[0]}=1$. For example, for the case of SU(3), multiplying
\suthreebd\ and \suthreebe\ by $3$ and subtracting it from this
rescaled solution yields one of our twisted solutions. Thus (for
SU(3)$/\Zop_3$ when 3 divides $k$) the solution \chafp\ is contained
in \suthreebd\ and \suthreebe\ (together with the twisted solutions)
but not vice versa. More generally, \chafp\ together with all of the
twisted solutions, will probably only generate a 
space of solutions to \basic\ which is of index 
gcd($d,M_{{\rm SU}(n)})$ in the complete space of solutions. It seems 
reasonably straightforward, given a specific group SU$(n)/\Zop_d$, to 
find the missing solutions (we shall find these for other general
classes in the next section), but we do not know a general expression
for them. For certain classes of theories we can however be more
specific. This will be described in the following section.

\newsec{The analysis for ${\rm SU}(n)/\Zop_d$, $d$ prime}

Until section 4.4,  we shall assume that $d$ is any prime dividing $n$.
If $d>2$ then the theory is always not pathological (\ie\ $(n+1)d'$
is even); the analysis of the previous section then shows that the
untwisted solution satisfies \basic\ modulo $M$, where $M$ is at least 
$M=M_{{\rm SU}(n)}/{\rm gcd}(d,M_{{\rm SU}(n)})$ and probably equal
to $M=M_{{\rm SU}(n)}$. In this section we shall give further
evidence that the untwisted solution actually solves \basic\ modulo 
$M=M_{{\rm SU}(n)}$. We shall also give a complete description of the
twisted charges (for all values of $k$) in this case; this is
described in section~4.2. 

The situation is different if $d=2$, since the theory may then be 
pathological (this happens precisely if $n/2$ is odd). If the theory
is not pathological the same arguments as for $d>2$ apply, and we can
give a complete description of the twisted charges and provide good
evidence that the untwisted solution satisfies \basic\ modulo 
$M_{{\rm SU}(n)}$. In the pathological case, the situation depends on
whether we are in Case A or Case B. In the former situation, the
charge group is either trivial or $K=\Zop_2\oplus \Zop_2$, while in
Case B it is either $\Zop_4$, $\Zop_2\oplus\Zop_2$, $\Zop_2$, or trivial. 
This is explained in section~4.3

In section~4.4 we briefly discuss the similarities and differences which
occur when $d$ is allowed to be composite. This helps to motivate the
conjectures we give in the concluding section.

To begin  let us collect some useful facts that will be needed
throughout this section.

\subsec{Some useful facts}

Let $\alpha$ be the highest power of $d$ dividing $n$. In
the following Claim, $d$ is assumed to be any prime.

\medskip\noindent{\bf Claim.} (a)  The greatest common divisor
gcd$({\rm dim}(\la))$, as $\la$ runs over all highest weights in  
$P_{+}^{k}$ with $d$ coprime to $t(\la)$, equals $d^{\alpha}$.

\noindent{(b)} For any fundamental weight 
$\Lambda_{\ell}$ of su($n$), 
\eqn\dimF{
{\rm dim}(\Lambda_{\ell})=\left\{
\eqalign{ {\rm dim}(\overline{\Lambda}_{\ell/d}) \qquad & 
{\rm if}\ d\ {\rm divides}\ \ell \cr
0\qquad \qquad &{\rm otherwise}}\right.\quad 
({\rm mod}\ d^{\alpha})\ ,}  
where dim$(\overline{\Lambda}_{{\ell/d}})$ is the dimension of the 
su($d'$) fundamental weight $\overline{\Lambda}_{\ell/d}$.

\noindent{(c)} Assume $d$ divides $k$. Then for any fundamental 
weight $\Lambda_{\ell}$ with $d$ dividing $\ell$,  and any 
$J^{d'}$-fixed points $\phi,\varphi\in P_{+}^{k}({\rm su}(n))$, 
\eqn\fusd{
N_{\Lambda_{\ell}\,\phi}{}^{\varphi}=
\overline{N}_{\overline{\Lambda}_{\ell/d}\,
\overline{\phi}}{}^{\overline{\varphi}}\,,}
where 
$\overline{\Lambda}_{\ell/d},\overline{\phi},\overline{\varphi}$ are 
the obvious weights in $P_{+}^{k/d}({\rm su}(d'))$.\medskip
 
The equality \fusd\ of fusion coefficients can be seen directly from the
Pieri rule. The bottom part of
\dimF\ follows from (a). For the top part of \dimF, write  
\eqn\dimFF{
{\rm 
dim}(\Lambda_{\ell})=\left({n\atop\ell}\right)
=\bigl(\prod_{i=0}^{\ell/d-1}{n-di\over\ell-di}
\bigr)\bigl(\prod_{{i=1\atop 
{\rm gcd}(d, i)=1}}^{\ell-1}{n-i\over \ell-i}\bigr)\,,}
where we have grouped together all terms with denominator and
numerator divisible by $d$ (first factor), and coprime to $d$ (second
factor). The first factor manifestly equals 
$\left({n/d\atop \ell/d}\right)$. It is divisible by
$d^{\alpha-\delta}$, where $d^{\delta}$ is the largest power of  
$d$ dividing $\ell$. The  numerator and denominator of each fraction 
${n-i\over \ell-i}$ in the second factor are manifestly congruent 
modulo $d^{{\rm min}(\alpha,\delta)}$. Thus the 
second factor is congruent to 1 modulo $d^{{\rm min}(\alpha,\delta)}$
(since the denominators are coprime to $d$, they can be inverted in
the ring  $\Zop_{d^{{\rm min}(\alpha,\delta)}}$). Putting this
together, we get (b). 

Finally, we turn to the proof of (a).
We know that the character ch$_\la$ (hence dimension dim$(\la)$) of any
weight with $t(\la)$ coprime to $d$, is a polynomial over $\Zop$  in
the characters
ch$_{\Lambda_\ell}$  (hence dimensions dim$(\Lambda_\ell$)) of the
fundamentals, where each term in that polynomial involves at least one
$\Lambda_\ell$ with $d$ coprime to $\ell$. So it suffices to
compute gcd(dim$(\Lambda_\ell)$) for $\ell$ coprime to $d$.
 First, this gcd must divide 
dim$(\Lambda_{1})=n$. If $p^m$ is the maximal power of a
prime $p\ne d$ dividing $n$, then 
dim$(\Lambda_{p^m})=\left({n\atop p^m} \right)$ will be coprime 
to $p$. On the other hand,  \dimFF\ 
shows $d^{\alpha}$ will divide any $\left({n\atop \ell}\right)$ for
$\ell$ coprime to $d$. This gives us part (a) of the claim.

\subsec{The case when $d$ is an odd prime}

Now suppose that $d$ is an odd prime dividing $n$, and denote as
before $d'=n/d$. In the previous section we found the general solution
for the case when $d$ and $k$ are coprime, since there are then no
fixed points. We now want to address the situation where $d$ and $k$
contain a common factor. Since $d$ is prime, this can only be the case
if $d$ divides $k$. In this case, a fixed point necessarily has order
$d$ with respect to $J^{d'}$. Note that the matrix $M[d']$ of
\simcurM\ will then be a simple current modular invariant
(\ie\ $(n+1)k d'$ is even), and \cond\ will be satisfied 
for $M=M_{{\rm SU}(n)}$.
 
The boundary states are parametrised by all $J^{d'}$-orbits $[\nu]$ 
in $P_+^k({\rm su} (n))$ where $\nu$ is not a fixed point of $J^{d'}$, 
together with exactly $d$ copies $[\varphi,i]_{1\le i\le d}$ for each 
fixed point $\varphi$. These fixed points $\varphi$ are in natural
bijection with the level $k/d$ weights $\overline{\varphi}$ of
su($d')$, as explained in section~1.1. The exponents consist of all
weights $\mu$ with $d$ dividing $t(\mu)$ (by \nalfp\ this includes all
$J^{d'}$-fixed points), where non-fixed points $\mu$ are counted
exactly once and the fixed points again come with multiplicity $d$: 
$(\phi,i)_{1\le i\le d}$. Although it may not be completely obvious, 
these two sets (namely boundary labels and exponents) always have the
same cardinality.  
 
Most entries of the $\psi$ matrix in \psinim\ are given by \psimu\ and 
\psinu. When both $\mu=\phi$ and $\nu=\varphi$ are fixed points, we
get 
\eqn\psiodfour{
\psi_{[\varphi,h]\,(\phi,j)}= {1\over d} \,
\left(S_{\varphi\phi}+\overline{S}_{\bar{\varphi}\bar{\phi}}(d\,
\delta^{h\, j} -1)\right)\,,}
where $\overline{S}_{\bar{\varphi}\bar{\phi}}$ is the $S$-matrix for 
su$(d')$ at level $k/d$. 

A technical observation makes it possible to proceed with an explicit
expression for the associated NIM-rep. {\it Fixed point factorisation}
\refs{\gw} explains how to express the su$(n)$ level $k$ $S$-matrix
entries $S_{\la\phi}$ involving $J^{d'}$-fixed points $\phi$ ($\la$ can
be arbitrary) in terms of the su$(d'$) level $k/d$ $S$-matrix entries
involving $\overline{\phi}$. With this observation, it is then 
possible to calculate the remaining NIM-rep coefficients 
$\N_{\la\, a}{}^b$ in terms of WZW fusion rules. Together with the
ones that were already given in \nimone, we can now calculate
\eqn\nimodtwo{ 
{\cal N}_{\Lambda_\ell\, [\phi,i]}{}^{[\varphi,j]}= \left\{
\eqalign{ 
 \delta^{ij} \,
\overline{N}_{\bar{\Lambda}_{\ell/d}
\bar{\phi}}{}^{\bar{\varphi}}\quad 
& {\rm if}\ d\ {\rm divides}\ \ell\cr
0 \qquad\qquad&{\rm otherwise,}}\right.}
where, as usual, the bar denotes su$(d')$ quantities at level $k/d$.
Analogous expressions can be found for other $\la$, but they are more
complicated.

As mentioned earlier, the NIM-rep is uniquely determined by its values 
for the fundamental weights $\la=\Lambda_\ell$, $1\le\ell\le n/2$, and
these (together with the simple currents) are also the most
useful. For convenience we also include the simple currents 
\eqn\nimodthree{
{\cal N}_{J^\ell 0\, [\phi,i]}{}^{[\varphi,j]}=
\delta_{J^\ell\phi\,\varphi}\, \delta^{ij}\,.}

With these preparations we can now determine the D-brane charges. 
Recall from the beginning of section~3 that both $M^u$ and $M^{t}$
divide $M_{{\rm SU}(n)}$. Furthermore, as explained above, 
\cond\ is automatically satisfied.

First let us determine the twisted charges, \ie\ the solutions  
$q^t_{[\nu,i]}$ to \basic\ with $q^t_{[0]}=0$. Then, because of
\twchnfp, $q^t_{[\nu]}=0$ for any $\nu$ that is not a fixed
point. Furthermore, we have \twchsum. By \nimone, we find that $M^t$
must divide dim$(\la)\,q^t_{[\phi,i]}$, whenever the fusion product of 
$\la$ and $\phi$ avoids all fixed points. By \scfus\ and \nalfp, this
will happen for example whenever $d$ does not divide $t(\la)$. Because
of claim (a) the gcd of the dimensions of these weights is 
$d^{\alpha}$. So $M^t$ must divide  
$d^{\gamma}={\rm gcd}(d^{\infty},n,M_{{\rm SU}(n)})$.  We will find
that equality works: 
\eqn\twchgp{
M^t=d^\gamma={\rm gcd}(d^\infty,n,M_{{\rm SU}(n)})\,.} 
 
Recall from section 1.1 that the fixed points of $P^k_+({\rm su}(n))$ are
in one-to-one correspondence with the weights in $P^{k/d}_+({\rm su}(d'))$.
In one direction, we project $\phi\in P^k_+({\rm su}(n))$ to $\overline{\phi}
\in P^{k/d}_+({\rm su}(d'))$, so that $\phi=(\overline{\phi},\overline{\phi},
\ldots,\overline{\phi})$ ($d$ times); for the inverse direction, we lift
$\nu\in P^{k/d}_+({\rm su}(d'))$ to $\widehat{\nu}=(\nu,\nu,\ldots,\nu)
\in P^k_+({\rm su}(n))$. For example, for SU$(6)/\Zop_3$, $\widehat{0}
=({k\over 3};0,{k\over 3},0,{k\over 3},0)$.

Considering \nimone\ and \nimodtwo, we find that for fixed $\phi$, the
values of  $q^t_{[\phi,i]}$ for different $i$ are independent,
except for the single relation \twchsum:
\eqn\twsum{
\sum_{i=1}^{d}q^t_{[\phi,i]}=0\qquad({\rm mod}\ d^{\gamma})\,.}
Choose any solution $q^{t}_{[\widehat{0},i]}$ to \twsum.  There are 
precisely $(d^{\gamma})^{d-1}$ of these, which we will show form the subgroup 
$\Zop_{d^{\gamma}}\oplus\cdots\oplus\Zop_{d^{\gamma}}$ of the charge 
group $K$. From Claim (c) we see that, for each $1\le \ell<d$, 
\eqn\verione{
{\rm dim}(\Lambda_{\ell 
d})\,q^{t}_{[\widehat{0},i]}=q^{t}_{[\widehat{\Lambda}_{\ell},i]}\,.}
Continuing recursively, we obtain all twisted
charges $q^{t}_{[\psi,i]}$. In fact, from Claim (b) we obtain the
remarkable formula, valid for all fixed points $\phi$: 
\eqn\twchod{
q^{t}_{[\phi,i]}
={\rm dim}(\overline{\phi})\,q^{t}_{[\widehat{0},i]}\,.}
\vskip4pt

Now let us determine the untwisted charges. We expect this to be a 
solution to \basic\ with $M=M_{{\rm SU}(n)}$ and  $q_{[0]}=1$.
In other words, we expect the total charge group to be 
\eqn\chgp{
K=\Zop_{M_{{\rm SU}(n)}}\oplus 
\Zop_{d^{\gamma}}\oplus\cdots\oplus\Zop_{d^{\gamma}}}
($d-1$ copies of $\Zop_{d^{\gamma}}$).
Unfortunately at present we can only prove the existence of this
solution in  special cases (which however exhaust most possibilities
for $n,k,d$). We can however describe a general ansatz for it, and
solve it for examples.

We know $q_{[\lambda]}= {\rm dim}(\lambda)$ for all non-fixed points 
$\la\in P_+^{k}$, so it suffices to study the fixed points $\phi$. 
When $d$ and $M_{{\rm SU}(n)}$ are coprime, \ie\ when the largest
power $d^{\beta}$  dividing $k+n$  obeys $d^{\beta}<n$, we get from
\chafp\ the unique general  
untwisted solution $q_{[\phi,i]}=d^{-1}{\rm dim}(\phi)$ where 
$d^{{-1}}$ is the inverse of $d$ mod $M_{{\rm SU}(n)}$. Thus it
suffices to consider $d$ dividing $M_{{\rm SU}(n)}$.

Let $\la$ be any weight with $t(\la)$ coprime to $d$. Then the tensor
product (hence fusion and NIM-rep \nimone) of $\la$ with any
$J^{d'}$-fixed point  $\phi$, cannot contain any $J^{d'}$-fixed
points, but will expand out into full $J^{d'}$-orbits. This means
\basic\ gives 
\eqn\dimphi{
{\rm dim}(\la)\,q_{[\phi,i]}={1\over d}{\rm dim}(\la)\,{\rm 
dim}(\phi)\quad \left({\rm mod}\ {M_{{\rm SU}(n)}\over d}\right)\,,}
where the ${1\over d}$ arises because only one orbit representative 
appears in \nimone. Doing this for all such $\la$ and using 
Claim (a), we obtain
\eqn\chfp{
q_{[\phi,i]}={1\over d}{\rm dim}(\phi)
+\ell_{i}(\phi) \,{M_{{\rm SU}(n)}\over d^{\gamma+1}}\,,}
where $0\le \ell_{i}(\phi)<d^{\gamma}$ are integers. Because this 
untwisted solution must have order dividing $M_{{\rm SU}(n)}$, the charges 
$q_{[\phi,i]}$ must be integers, and thus
\eqn\twint{
{M_{{\rm SU}(n)}\over d^\gamma}\, 
\ell_{i}(\phi)=-{\rm dim}(\phi)\qquad ({\rm mod}\ d)\,.}
(This is also the reason we took $d^{\gamma+1}$ in \chfp\ rather than
$d^{\alpha+1}$.) Of course these integers must also obey the condition
\eqn\sumell{
\sum_{i=1}^{d}\ell_{i}(\phi)=0\,.}
These conditions cannot uniquely determine the $\ell_i(\phi)$, as we 
can always add to them $M_{{\rm SU}(n)}/d^{\gamma}$ times any twisted
solution. 

We do not have a proof that in general these integers $\ell_i(\phi)$ can be  
found. However, the untwisted solution {\it must} look like \chfp, for 
integers $\ell_i(\phi)$ satisfying \twint\ and \sumell, if that 
solution is to have order $M_{{\rm SU}(n)}$ (the maximal possible). 
We also know that \chafp\ will work mod $M_{{\rm SU}(n)}/d$ and
that it has $q_{[0]}=1$. So the only question is whether this
solution can be lifted to $M_{{\rm SU}(n)}$, modulo twisted solutions. 


While we cannot prove this at present, we can give various pieces of
evidence in favour of \chfp. First of all, we note that it requires
$d$ to divide dim$(\phi)$ whenever $d^{\alpha+1}$ divides $M_{{\rm SU}(n)}$. 
This is indeed the case as follows from \basicsun, \cond, and Claim (a): if
$d^{\alpha+1}$ divides $M_{{\rm SU}(n)}$ then for any fixed   
point $\phi$
\eqn\calc{
d\,{\rm dim}(\phi)
=\sum_{\nu}N_{\Lambda_{1}\,\phi}{}^{\nu}{\rm dim}(\nu)
= d\sum_{[\nu]}N_{\Lambda_{1}\,\phi}{}^{\nu}=0\qquad({\rm mod}\ 
d^{\alpha+1})\,.}
 
Secondly, for the case where $d=n$ we can construct a solution with  
$M=M_{{\rm SU}(n)}$, generalising the solution 
\suthreebe\ for SU$(3)/\Zop_3$ and proving \chgp\ holds for prime $n$. 
To this end we put, for any fixed point $\phi\in P_+^k$,
\eqn\sunbe{
q_{[\phi,i]}={1\over d}{\rm dim}(\phi+\Lambda_1)
+\ell_{i}(\phi)\,{M\over d}\,,}
where the integers $0\le\ell_{i}(\phi)<d$ are chosen so that \chargesum\ is 
satisfied. It is easy to verify that $d$ does divide
dim$(\phi+\Lambda_1)$. By Claim (b), $d$ will divide each
dim$(\Lambda_i)$ here and so \basic\ will automatically be
satisfied. Incidentally, the only significance of the number 
${\rm dim}(\phi+\Lambda_1)$ in \sunbe\ or \suthreebe\ is that it is
divisible by $d$ and it is congruent to dim$(\phi)$ mod 
$M_{{\rm SU}(n)}/d$.
 
Finally, we consider the example SU$(6)/\Zop_{3}$. It suffices to consider
the situation when $3$ divides $M=M_{{\rm SU}(6)}$, where
$M_{{\rm SU}(6)}=(k+6)/(2^{i}3)$ and $2^i={\rm gcd}(4,k+6)$. The fixed
points are of the form $\phi=(\phi_1,\phi_0,\phi_1,\phi_0,\phi_1)$,
where $\phi_0=k/3-\phi_1$. Write $\phi_1'$ for $\phi_{1}+1$ and $k'$
for $k+6$. Then the Weyl dimension formula reads 
\eqn\susix{\eqalign{
{\rm dim}(\phi)&\,={\phi'_{1}{}^{3}
(k'/3-\phi'_{1})^{2}(k'/3)^{4}(k'/3+\phi'_{1})^{2}
(2k'/3-\phi'_{1})(2k'/3)^{2}(2k'/3+\phi'_{1})\over 
2^{4}3^{3}4^{2}5}\cr &\,=
(k'/3)^{6}{\phi'_{1}{}^{3}((k'/3)^{2}-\phi'_{1}{}^{2})^{2} 
((2k'/3)^{2}-\phi'_{1}{}^{2})\over 2^{6}3^{3}5}\,.}}
From this we obtain that dim$(\phi)$ is always a multiple of $3M$ (in
fact $9M$). The ansatz \chfp\ then tells us that 
$q_{[\phi,i]}=\ell'_{i}(\phi)\,M/3$ for some integers 
$\ell'_{i}(\phi)$, and adding an appropriate twisted solution \twchod, 
we know we can choose $q_{[\phi,i]}=0$. Indeed, an easy calculation  
verifies that the sum 
$\sum_{[\nu]}N_{\Lambda_{i}\phi}{}^{\nu}{\rm dim}(\nu)$ over all
$J^{2}$-orbits of non-fixed points is divisible by $3$ for all fixed 
points $\phi$ and fundamentals $\Lambda_i$ (even though $3$ does not 
divide all  dim$(\nu)$ with $N_{\Lambda_{i}\phi}{}^{\nu}\ne 0$). Thus
the general untwisted solution, for SU$(6)/\Zop_{3}$ when $3$ divides
$M$ (\ie\ $k=3$ mod 9), has $q_{[\phi,i]}=0$ for each fixed point
$\phi$. The charge group is therefore indeed 
$K=\Zop_{M_{{\rm SU}(6)}}\oplus\Zop_3\oplus\Zop_{3}$, as expected from
\chgp.

\subsec{The case when $d=2$}

When $d$ is even, the theory can be pathological, and in particular,
Case B can arise.
As usual we compute the NIM-rep coefficients from the $\psi$ matrix. 
Its only entries that have not yet been given are those between fixed
points: 
\eqn\psitwo{
\psi_{[\phi,i]\,[\varphi,j]}={1\over 
\sqrt{2\,{\rm mult}(\varphi)}}\left(S_{\phi\,\varphi}
+(-1)^{{i+j}}e^{-3\pi {\rm i} kn/16}\overline{S}_{
\overline{\phi}\overline{\varphi}}\right)\,.}
The strange relative phase $e^{-3\pi {\rm i} kn/16}$ is the modular
anomaly shift and plays no role in \psinim\ or elsewhere in this paper.
From \psitwo\ we readily compute from \psinim\ the remaining entries of $\N$:
\eqn\nimtwo{
\N_{\Lambda_{m}\,[\phi,i]}{}^{[\varphi,j]}=\left\{
\eqalign{0\qquad \quad& {\rm if}\ m\ {\rm is\ odd}\cr 
\delta^{ij}\, \overline{N}_{\overline{\Lambda}_{m/2}\overline{\phi}}
{}^{\overline{\varphi}}\quad & {\rm otherwise.}}\right.} 
Equations \psitwo\ and \nimtwo\ hold for both Case A and Case B. 
In order to show in Case B that \nimtwo\ arises from \psitwo\ requires
\nalfp, which implies $S_{\phi\varphi}=0$ for any order 2 fixed points 
$\phi,\varphi$. As we have seen before, in Case A the simple current
$\N_{J^{d'}0}$ is the identity, but in Case B it satisfies
\eqn\sctwo{
\N_{J^{d'}0\,[\phi,i]}{}^{[\varphi,j]}
=\delta^{(2)}_{j,i+1}\,\delta_{\phi,\varphi}\,,}
whenever $\phi,\varphi$ are order-two fixed points. Furthermore, 
$\N_{J^{d'}0\,[\nu]}{}^{[\nu']}=\delta_{[\nu],[\nu']}$, and all other
entries are $0$.  

By \dimJtwo, we find that dim$(J^{n/2}0)=(-1)^{n/2}$ 
(mod $M_{{\rm SU}(n)}$). When $n/2$ is even, the theory is not
pathological, and the analysis and formulae are essentially identical
to that in the previous subsection. In particular the matrix $M[d']$
is then a modular invariant for any value of $k$. We can construct the
twisted solution \twchod\ as before, and the charge group is therefore
very plausibly \chgp\ with $M=M_{{\rm SU}(n)}$ and
$M^{t}=2^{\gamma}$. We can prove that this is the correct answer
whenever $M_{{\rm SU}(n)}$ is odd since then the arguments of
section~3 apply. [The construction of the twisted charges works mod
$M^t$ in any case, since the relevant arguments of section~4.2 apply
here as well.] 

As before, we cannot give a proof in the other cases, but we can at
least illustrate our claim with an example. To this end consider 
SU(4)$/\Zop_2$ for the case when $M_{{\rm SU}(4)}$ is even; this can
only occur when 4 divides  $k$, in which case 
$M\equiv M_{{\rm SU}(4)}=(k+4)/(2\cdot 3^i)$, where $i=1,0$ depending
on whether $k+4$ is divisible by $3$ or not. The $J^2$ fixed points
are then  $\phi^{a}=(a,k/2-a,a)$ for $0\le a\le k/2$. The twisted
solution \twchod\ becomes 
$q^{t}_{[\phi^{a},i]}=(-1)^{i}(a+1)q^{t}_{[\phi^{0},2]}$ where
$i=1,2$. This defines a solution mod $2^\gamma$, where $2^\gamma=4$ if
$M/2$ is even, and $2^\gamma=2$ if $M/2$ is odd. The untwisted
solution is given, if $M/2$ is odd, by $q_{[\phi^{2a},i]}= (a+i)M/2$
for $a=0,1,\ldots,k/4$ and $i=1,2$, and $q_{[\phi^{b},j]}=0$ 
otherwise. When 4 divides $M$, $q_{[\phi^a,i]}=M/2$ for $a=1$ mod 4
and $i=1,2$, and $q_{[\phi^{b},j]}=0$ otherwise. It is not difficult
to check that these charges work, and that this gives rise to the
charge group \chgp\ with $M^u=M_{{\rm SU}(4)}$ and $M^{t}=2^{\gamma}$.  
\vskip8pt

This leaves us with analysing the pathological case, \ie\ the
situation when $n/2$ is odd. As mentioned before, in this case 
the matrix $M[d']$ will only be a modular invariant if $k$ is
even, so let us assume $k$ even in the following. (If $k$ is odd,
there are no fixed points, and the analysis of section~3.2 already
gives the complete solution.) Furthermore, \cond\ implies that $M=2$
(if $M_{{\rm SU}(n)}$ is even) or $M=1$ (when it is odd). As the
example of SO$(3)$ demonstrates, the charge group will depend on
whether we are in Case A (which arises if $k/2$ is even) or in Case B
(if $k/2$ is odd). 

As we have seen before in section~3.1, in Case A the charge group will
be a direct sum of $\Zop_2$'s. If $M_{{\rm SU}(n)}$ is odd then 
$\Zop_2$ is not possible (the order of any charge group must divide
$M_{{\rm SU}(n)}$), and the charge group is necessarily trivial.  
This will happen whenever the exact power of $2$ dividing $k+n$, is
less than $n$. The only interesting case is therefore 
$M_{{\rm SU}(n)}$ even, in which case the arguments of section~3.1
imply that $M^u=M^t=2$. The construction of the twisted solution 
\twchod\ still works, and thus the twisted solution will give one
factor of $\Zop_2$ to \Kgen. 
We conjecture, as in the previous case, that there
always exists an untwisted solution that works mod $2$, \ie\ that it
is non-trivial. Thus we conjecture that the complete charge group is  
$\Zop_2\oplus\Zop_2$ if $M_{{\rm SU}(n)}$ is even.
\vskip4pt

In Case B, the twisted solution \twchod\ still works mod $M^t$, where 
\eqn\possBcgp{
M^t=M={\rm gcd}(2,M_{{\rm SU}(n)})\in\{1,2\}\,.}
Far less trivial is however the nature of the untwisted solution. For 
example, the case of SO$(3)$ with $k$ not divisible by $4$ shows that
the untwisted solution may have order $M^u=4$, in which case it includes
the above twisted solution. On the other hand, the example of
SU$(6)/\Zop_2$ (which falls into Case B when $k/2$ is odd) shows that 
the untwisted solution may not exist altogether: when 
$k=2$ mod 16, $M={\rm gcd}(2,M_{{\rm SU}(n)})=2$, so we would
expect an untwisted  solution to \basic\ with $q_{[0]}=1$ and $M=2$,
and hence $q_{[\nu]}={\rm dim}(\nu)$. However, taking
$\lambda=\Lambda_1$ and $a=[(0,0,{k\over 2},0,0),i]$, we find that
$q_a\in\Zop+{1\over 2}$ and so the desired untwisted solution would
have to have order $4$. This contradicts the observation made at the
beginning of section~3, that the order $M^u$ must divide 
$M_{{\rm SU}(n)}$. Thus we conclude that the untwisted solution must
be trivial, \ie\ that it has order $M^u=1$. The total charge
group of this theory is then $K=\Zop_2$. [Incidentally, 
this behaviour is special to these values of $k$: when $k=6$ mod $8$,
$M=M^t=1$ and the charge group is trivial. Also, when $k=10$ mod $16$,
the charge group is $K=\Zop_2\oplus\Zop_2$, reflecting
the separate existence of an untwisted and a twisted solution.]

We conjecture that the total charge group in Case B is (i) trivial if
$M_{{\rm SU}(n)}$ is odd; is (ii) $\Zop_2\oplus\Zop_2$ if 
$M_{{\rm SU}(n)}$ is even and the dimension of
$\dim(\widehat{0}+\Lambda_1)$ is even; is (iii) $\Zop_4$ if 
$M_{{\rm SU}(n)}$ is divisible by $4$ and the dimension of
$\dim(\widehat{0}+\Lambda_1)$ is odd; and (iv) is $\Zop_2$ if 
$M_{{\rm SU}(n)}$ is even but not divisible by $4$ and the dimension
of $\dim(\widehat{0}+\Lambda_1)$ is odd. Here $\widehat{0}+\Lambda_1$ is a
representative weight of the unique $J^{d'}$-orbit that appears in the 
fusion of $\Lambda_1$ with the fixed point $\widehat{0}=
(k/d;0,\ldots,0,k/d,0,\ldots,0,\ldots,k/d,0,\ldots,0)$.

\subsec{Some remarks about composite $d$}

We believe that the behaviour of solutions to \basic\ in the special
case where $d$ is prime, is quite representative of what happens in
general (\ie\ when $d$ is composite). In particular, simple formulae
such as \nimodtwo\ continue to hold \gaga\ (though generally with sums
over simple currents) for general SU$(n)/\Zop_d$, and so the analysis
of the previous subsections can largely follow through. 

Equation \twchod\ says that all (twisted) fixed point charges can be
recovered from those of the basic one $\widehat{0}$, which corresponds
to the vacuum in $P_+^{k/d}({\rm su}(d'))$. This continues to hold in
all composite examples we have considered, and is the basis for our
conjectures in the next section. 

There is one new phenomenon which can happen for composite
$d$. Consider for example SU$(9)/\Zop_9$ at level $18$, for which
$M_{{\rm SU}(9)}=9$. Then \basic\ with $\la=\Lambda_3$ and
$a=\widehat{0}$ implies
\eqn\fone{
84\, q^t_{[\widehat{0},i]} = q^t_{[\widehat{(1;3,2)},i]}\,.}
The fixed point on the left has order 9, while the fixed point on the 
right has order 3. Thus if we replace $i$ by $i+3$ on the left hand
side, we must get the same right hand side. Hence we obtain the
constraint that 
$$
84\, (q^t_{[fp9,i]} - q^t_{[fp9,i+3]}) = 0\qquad   ({\rm mod}\ M^t) \,.
$$
Since $M^t=9$ here, this constitutes a nontrivial constraint on the
twisted charges. Together with \twchsum, this reduces the contribution
to $K$ of the twisted solutions from the usual $\Zop_9^9$ to 
$\Zop_9^3\oplus \Zop_3^5$. More generally, we get the constraint that
$M^t$ must divide 
$({n\atop \ell})(q^t_{[\widehat{0},i]}- q^t_{[\widehat{0},i+\ell]})$
for all $\ell$.

\newsec{Conclusions}

In this paper we have analysed the charges of D-branes for string
theory with target space SU$(n)/\Zop_d$, where $d>1$ divides
$n$. Based on our results one may conjecture that there are
essentially two (or maybe three) different cases that need to be
distinguished:  
\vskip4pt

\noindent {\bf (1) If $(n+1)n/d$ is even (the non-pathological case)},
then the charge group is a subgroup of 
\eqn\finalr{
K = \Zop_{M_{{\rm SU}(n)}} \oplus \left(\Zop_{M^t}\right)^{f-1}\,,}
where $f={\rm gcd}(d,k)$ and where 
\eqn\finalMt
{M^t={\rm gcd}(d^\infty,n,M_{{\rm SU}(n)})\,.}
[Note that if $k$ is coprime to $d$, the charge group is simply   
$K=\Zop_{M_{{\rm SU}(n)}}$. In all examples we have studied, $K$
always contained $\Zop_{M_{{\rm SU}(n)}}$, but we know of examples
--- see section 4.4 --- where the twisted solutions only gave a proper 
subgroup of  $(\Zop_{M^t})^{f-1}$.]
\smallskip

\noindent {\bf (2) If $(n+1)n/d$ is odd (the pathological case)}, then
the charge group is `pathologically small', \ie\ it is bounded as
$k\rightarrow\infty$. More precisely, when $k/f$ is even 
(`{\bf Case A}') the charge group is either trivial (if 
$M_{{\rm SU}(n)}$ is odd), or it is a direct sum of at most $f$
copies of the cyclic group $\Zop_2$. 

When $k/f$ is odd (`{\bf Case B}'), the charge group will
be a subgroup of $\Zop_M\oplus(\Zop_{M^t})^{f-1}$ or
$\Zop_{MM^t}\oplus(\Zop_{M^t})^{f-2}$, 
where $M={\rm gcd}(2,M_{{\rm SU}(n)})$, and $f$ and $M^t$ are given as 
above.
\vskip4pt

Our conjectures are the simplest statements we have found which are
consistent with all of the examples and results we know.
We have not been able to prove these claims in the above generality,
but we have been able to prove them for most cases (for example, when
$d$ is prime, when $k$ is coprime to $d$, {\it etc}). We have also
illustrated these results with two examples:
SU$(3)/\Zop_3$ is the archetypal example for case (1), while 
SO$(3)=$ SU$(2)/\Zop_2$ is a good example for case (2). As explained in section
4.4, we expect that a solution to \basic\ is uniquely determined from the
vacuum charge $q_{[0]}$, together with the $f$ charges
$q_{[\widehat{0},i]}$ associated to the order-$f$ fixed point
$\widehat{0}=\sum_{\ell=0}^{f-1}{k\over f}\Lambda_{jn/f}$. 
These charges are not independent --- in particular there is
(3.8), and therefore the ambiguity in the construction of the
untwisted solution is at most $(\Zop_{M^t})^{f-1}$. Also, if $d$ is not
prime, there can be additional constraints (see section 4.4).

Some of these observations will generalise to other WZW models. For
example, SO$(2n+1)=$B$_n/\Zop_2$ will behave like case (2) above for
every $n$ and $k$. 

It is intriguing that for the quotient groups of SU$(n)$, the
pathological case (2) appears precisely for those theories where the
quantisation condition of \refs{\fgka} is $k\in 2\Zop$. This suggests
that one should be able to understand the pathological behaviour of
the D-brane charges also from a geometrical point of view. At any
rate, one should expect that the above results should agree with the
charge groups that can be determined by a K-theory analysis, and it
would be very interesting to check this. The behaviour of the D-brane 
charges for non-simply connected groups is far more subtle than in the
simply connected case, and this should also be reflected in the
K-theory analysis. 

It is believed that the K-theory analysis should apply to the
supersymmetric version of this model. As is well known, the
supersymmetric theory can be rewritten in terms of the bosonic WZW
model (at a shifted level) together with $\dim(\bar{{\frak g}})$ free
fermions. In this paper we have only analysed the bosonic part in
detail; since we have only considered D-branes that preserve the full
affine symmetry, it is always possible to choose boundary conditions
for the free fermions so that the combined D-brane preserve the full
supersymmetry. Thus the D-branes we have discussed here always
correspond to supersymmetric branes. 

There is very good evidence that the D-brane charges must satisfy
\basic, but it is not so obvious whether this is the only constraint
that restricts them. In particular, it is conceivable that
there are other constraints that remove some of the ambiguities that
are described by the twisted solutions. 

Finally, we have only discussed the maximally symmetric D-branes for
these theories, but it is clear that there are also `twisted' D-branes
that only preserve the affine symmetry up to an outer
automorphism. It should be possible to determine their charges using
similar methods. Given the situation for the simply connected case,
one should also expect that the full charge lattice will not be
generated by these D-branes alone. It would therefore be interesting
(as in the simply connected case) to understand the structure of the
remaining charges.

\vskip 1cm

\centerline{{\bf Acknowledgements}}\par\noindent

\noindent We thank Stefan Fredenhagen for conversations and
correspondences. This paper was begun while both of us were visiting
BIRS, and we are grateful for their generous hospitality. 
Parts were also written while TG was visiting IHES and University of
Wales Swansea, and he warmly thanks both for their hospitality. 
TG's research is supported in part by NSERC.

\vskip1.5cm

\noindent {{\bf Note added in proof.}} The K-theory 
calculation for SO(3) has recently appeared \refs{\bsn}; they recover
our charge groups, as expected, but they also find evidence for a
second supersymmetric conformal field theory associated to SO(3). A
proposed construction of this novel SO(3) model has been given in
\refs{\fre}, where it is shown to recover the second family of
possible SO(3) charge groups of \refs{\bsn}. 
\vskip1.5cm

\listrefs

\bye